\documentclass[12pt,twocolumn]{aastex631}
\usepackage{graphics}
\usepackage{color}
\usepackage{topcapt}
\usepackage{amsmath}
\usepackage{empheq} 
\newcommand{\ltsimeq}{\la}

\newcommand{\msun}{M$_{\odot}$}
\newcommand{\mstar}{$M_*$}

\newcommand{\hii}{H{\sc ii}}

\newcommand{\tten}{$\tau_{10}$}
\newcommand{\ttwentyfive}{$\tau_{25}$}
\newcommand{\tfifty}{$\tau_{50}$}
\newcommand{\tninety}{$\tau_{90}$}

\shortauthors{McQuinn et al.}
\shorttitle{The Star Formation History of WLM with JWST}

\begin{document}

\title{The JWST Resolved Stellar Populations Early Release Science Program IV: The Star Formation History of the Local Group Galaxy WLM}
\correspondingauthor{Kristen.~B.~W. McQuinn}
\email{kristen.mcquinn@rutgers.edu}

\author[0000-0001-5538-2614]{Kristen.~B.~W. McQuinn}
\affiliation{Rutgers University, Department of Physics and Astronomy, 136 Frelinghuysen Road, Piscataway, NJ 08854, USA}

\author[0000-0002-8092-2077]{Max~J.~B. Newman}
\affiliation{Rutgers University, Department of Physics and Astronomy, 136 Frelinghuysen Road, Piscataway, NJ 08854, USA}

\author[0000-0002-1445-4877]{Alessandro Savino}
\affiliation{Department of Astronomy, University of California, Berkeley, CA 94720, USA}

\author[0000-0001-8416-4093]{Andrew E. Dolphin}
\affiliation{Raytheon Technologies, 1151 E. Hermans Road, Tucson, AZ 85756, USA}
\affiliation{Steward Observatory, University of Arizona, 933 N. Cherry Avenue, Tucson, AZ 85719, USA}

\author[0000-0002-6442-6030]{Daniel R. Weisz}
\affiliation{Department of Astronomy, University of California, Berkeley, CA 94720, USA}

\author[0000-0002-7502-0597]{Benjamin F. Williams}
\affiliation{Department of Astronomy, University of Washington, Box 351580, U.W., Seattle, WA 98195-1580, USA}

\author[0000-0003-4850-9589]{Martha L. Boyer}
\affiliation{Space Telescope Science Institute, 3700 San Martin Drive, Baltimore, MD 21218, USA}

\author[0000-0002-2970-7435]{Roger E. Cohen}
\affiliation{Rutgers University, Department of Physics and Astronomy, 136 Frelinghuysen Road, Piscataway, NJ 08854, USA}

\author[0000-0001-6464-3257]{Matteo Correnti}
\affiliation{INAF Osservatorio Astronomico di Roma, Via Frascati 33, 00078, Monteporzio Catone, Rome, Italy}
\affiliation{ASI-Space Science Data Center, Via del Politecnico, I-00133, Rome, Italy}

\author[0000-0003-0303-3855]{Andrew A. Cole}
\affiliation{School of Natural Sciences, University of Tasmania, Private Bag 37, Hobart, Tasmania 7001, Australia}

\author[0000-0002-7007-9725]{Marla C. Geha}
\affiliation{Department of Astronomy, Yale University, New Haven, CT 06520, USA}

\author[0000-0002-5581-2896]{Mario Gennaro}
\affiliation{Space Telescope Science Institute, 3700 San Martin Drive, Baltimore, MD 21218, USA}
\affiliation{The William H. Miller {\sc III} Department of Physics \& Astronomy, Bloomberg Center for Physics and Astronomy, Johns Hopkins University, 3400 N. Charles Street, Baltimore, MD 21218, USA}

\author[0000-0002-3204-1742]{Nitya Kallivayalil}
\affiliation{Department of Astronomy, The University of Virginia, 530 McCormick Road, Charlottesville, VA 22904, USA}

\author[0000-0002-4378-8534]{Karin M. Sandstrom}
\affiliation{Department of Astronomy \& Astrophysics, University of California San Diego, 9500 Gilman Drive, La Jolla, CA 92093, USA}

\author[0000-0003-0605-8732]{Evan D. Skillman}
\affiliation{University of Minnesota, Minnesota Institute for Astrophysics, School of Physics and Astronomy, 116 Church Street, S.E., Minneapolis,
MN 55455, USA}

\author[0000-0003-2861-3995]{Jay Anderson}
\affiliation{Space Telescope Science Institute, 3700 San Martin Drive, Baltimore, MD 21218, USA}

\author[0000-0002-5480-5686]{Alberto Bolatto}
\affiliation{Department of Astronomy, University of Maryland, College Park, MD 20742, USA}
\affiliation{Joint Space-Science Institute, University of Maryland, College Park, MD 20742, USA}

\author[0000-0002-9604-343X]{Michael Boylan-Kolchin}
\affiliation{Department of Astronomy, The University of Texas at Austin, 2515 Speedway, Stop C1400, Austin, TX, 787125, USA}

\author[0000-0001-9061-1697]{Christopher T. Garling}
\affiliation{Department of Astronomy, The University of Virginia, 530 McCormick Road, Charlottesville, VA 22904, USA}

\author[0000-0003-0394-8377]{Karoline M. Gilbert}
\affiliation{Space Telescope Science Institute, 3700 San Martin Dr., Baltimore, MD 21218, USA}
\affiliation{The William H. Miller III Department of Physics \& Astronomy, Bloomberg Center for Physics and Astronomy, Johns Hopkins University, 3400 N. Charles Street, Baltimore, MD 21218, USA}

\author[0000-0002-6301-3269]{L\'eo Girardi}
\affiliation{Padova Astronomical Observatory, Vicolo dell'Osservatorio 5, Padova, Italy}

\author[0000-0001-9690-4159]{Jason S. Kalirai}
\affiliation{John Hopkins Applied Physics Laboratory, 11100 Johns Hopkins Road, Laurel, MD 20723, USA}

\author[0000-0002-7503-5078]{Alessandro Mazzi}
\affiliation{Dipartimento di Fisica e Astronomia Galileo Galilei, Universit\`a di Padova, Vicolo dell’Osservatorio 3, I-35122 Padova, Italy}

\author[0000-0002-9300-7409]{Giada Pastorelli}
\affiliation{Padova Astronomical Observatory, Vicolo dell'Osservatorio 5, Padova, Italy}

\author[0000-0002-3188-2718]{Hannah Richstein}
\affiliation{Department of Astronomy, The University of Virginia, 530 McCormick Road, Charlottesville, VA 22904, USA}

\author[0000-0003-1634-4644]{Jack T. Warfield}
\affiliation{Department of Astronomy, The University of Virginia, 530 McCormick Road, Charlottesville, VA 22904, USA}

\begin{abstract}
 We present the first star formation history (SFH) and age-metallicity relation (AMR) derived from resolved stellar populations imaged with the JWST NIRCam instrument. The target is the Local Group star-forming galaxy WLM at 970 kpc. The depth of the color-magnitude diagram (CMD) reaches below the oldest main sequence turn-off with a SNR$=10$ at M$_{F090W}=+4.6$ mag; this is the deepest CMD for any galaxy that is not a satellite of the Milky Way. We use Hubble Space Telescope (HST) optical imaging that overlaps with the NIRCam observations to directly evaluate the SFHs derived based on data from the two great observatories. The JWST and HST-based SFHs are in excellent agreement. We use the metallicity distribution function measured from stellar spectra to confirm the trends in the AMRs based on the JWST data. Together, these results confirm the efficacy of recovering a SFH and AMR with the NIRCam F090W$-$F150W filter combination and provide validation of the sensitivity and accuracy of stellar evolution libraries in the near-infrared relative to the optical for SFH recovery work. From the JWST data, WLM shows an early onset to star formation, followed by an extended pause post-reionization before star formation re-ignites, which is qualitatively similar to what has been observed in the isolated galaxies Leo~A and Aquarius. Quantitatively, 15\% of the stellar mass formed in the first Gyr, while only 10\% formed over the next $\sim$5 Gyr; the stellar mass then rapidly doubled in $\sim2.5$ Gyr, followed by constant star formation over the last $\sim5$ Gyr.
 \end{abstract} 

\section{Introduction}\label{sec:intro}
The James Webb Space Telescope (JWST) is capable of imaging stars in nearby galaxies to deeper photometric depths, at higher angular resolution, and with greater photometric precision than ever before possible. Such deep imaging is information-rich, and can provide robust constraints on the stellar mass assembly of galaxies over cosmic timescales. Specifically, color-magnitude diagrams (CMDs) from high-fidelity photometry of stars, including faint, old, low-mass stars (i.e., stars below the oldest main sequence turn-off; oMSTO) enable the reconstruction of robust star formation histories (SFHs) with high temporal resolution at all ages. These SFHs help elucidate the fundamental process of star formation that drives much of the cosmic baryon cycle and provide key benchmarks for testing galaxy formation and evolution theories. JWST therefore will transform our understanding of the growth of low-mass galaxies via archeological studies of resolved stars in nearby galaxies. 

\subsection{HST's SFH Legacy}
High temporal resolution SFHs from CMD-fitting techniques were originally made possible because of the Hubble Space Telescope's (HST's) sensitivity and image quality, and the ability to perform precision point-spread function (PSF) photometry of resolved stars down to faint magnitudes \citep[e.g.,][]{Dolphin2000, Dolphin2002, Tolstoy2009}. These SFHs have been instrumental in shaping our understanding of how galaxies build their stellar mass as a function of time. The scientific insights gained with the SFHs from HST data are quite diverse and, in large part, are determined by the photometric depth of a given data set. 

For example, SFHs have been derived for many galaxies within the Local Group (i.e., very nearby systems) from photometric catalogs, with the majority of those catalogs reaching depths below the oMSTO. The oMSTO feature helps to break the age-metallicity degeneracy for older-age stars in a CMD, thereby facilitating robust SFH recovery with temporal resolution as fine as log($\delta$t) $= 0.05$ at the oldest lookback times. These SFHs characterize the properties of satellite galaxies around the Milky Way and M31, including investigating quenching timescales and possible host-galaxy dependencies \citep[e.g.,][]{Monelli2010, Weisz2014a, Skillman2017, Weisz2019, Sacchi2021, McQuinn2023a, McQuinn2023b, Savino2023}. SFHs of the Large and Small Magellanic Clouds from similarly deep HST data and ground-based data have revealed synchronized bursts of star formation that are likely interaction-driven \citep{Harris2009, Weisz2013, Massana2022}. SFHs are currently being derived from $>100$ HST pointings across the Clouds in an effort to quantify the radial variations in their histories (Roger Cohen et al.\ in preparation). SFHs have also constrained the early evolution of a handful of gas-rich galaxies located within the Local Group at farther distances from the MW and M31 than the satellite systems, revealing intriguing differences on when these systems began to form their stellar mass in earnest \citep[e.g.,][]{Cole2007, Hidalgo2011, Skillman2014, Cole2014, Albers2019}. 

\begin{table}
\begin{center}
\caption{Properties and Observational Details of WLM}
\label{tab:properties}
\end{center}
\begin{center}
\vspace{-30pt}
\setlength{\tabcolsep}{0.5pt}
\begin{tabular}{lr}
\hline 
\hline 
\multicolumn{2}{c}{WLM Properties} \\
\hline
$M_V$ (mag)		& $-$14.2 (1) \\
$r_h$ (\arcmin)	& 7.8 (1) \\
E(B$-$V) (mag)	& 0.03  (2) \\
Distance (kpc)	& 968$^{+41}_{-40}$ (3)\\
$\mu$	(mag)	& 24.93$\pm$0.09 (3) \\
Supergiant stars $\langle$[Z]$\rangle$ & $-0.87\pm0.06$ (4) \\
Gas-phase 12$+$log(O/H)	& 7.83$\pm0.06$ (5) \\ 
RGB stars $\langle$[Fe/H]$\rangle$ & $-$1.14; $\sigma=0.33$ (6)\\

\hline
\multicolumn{2}{c}{JWST NIRCam Observations} \\
\hline
PID			    & ERS-1334\\
RA (J2000)		& 00:01:57.429 \\
Dec (J200)		&  $-$15:28:52.49 \\
F090W (s)		& 30492 \\ 
F150W (s)		& 23707 \\	
F090W 50\% Completeness (mag)	& 29.23 ($+$4.30)\\
F150W 50\% Completeness (mag)	& 27.89 ($+$2.96) \\
log(\mstar/\msun), Full FoV, PARSEC & $7.12\pm0.01$ \\
log(\mstar/\msun), Full FoV, BaSTI  & $7.13\pm0.01$ \\
log(\mstar/\msun), ACS FoV, PARSEC & $6.54\pm0.01$ \\
log(\mstar/\msun), ACS FoV, BaSTI & $6.58\pm0.01$ \\
log(\mstar/\msun), Joint, PARSEC & $6.64\pm0.01$ \\
log(\mstar/\msun), Joint, BaSTI & $6.65\pm0.01$ \\
\tten\ PARSEC; BaSTI (Gyr)  &  $12.61^{+0.06}_{-0.06}$; $13.12^{+0.02}_{-0.03}$ \\
\ttwentyfive\ PARSEC; BaSTI (Gyr)  &  $7.63^{+0.05}_{-0.03}$;  $7.87^{+0.06}_{-0.03}$\\
\tfifty\ PARSEC; BaSTI (Gyr)   &  $5.08^{+0.05}_{-0.01}$; $5.17^{+0.02}_{-0.02}$\\
\tninety\ PARSEC; BaSTI (Gyr)  & $1.18^{+0.01}_{-0.01}$; $1.11^{+0.01}_{-0.01}$\\
\hline
\multicolumn{2}{c}{HST ACS Observations} \\
\hline
PID			    & HST-GO-13768 \\
RA (J2000)		& 00:01:57.319 \\
Dec (J200)		& $-$15:31:22.52 \\
F475W (s)		& 27360 \\
F814W (s)		& 34050 \\
F475W 50\% Completeness (mag) & 28.28 ($+$3.35)\\
F814W 50\% Completeness (mag) & 27.20 ($+$2.27) \\
log(\mstar/\msun), NIRCam FoV, PARSEC & $6.60\pm0.01$\\
log(\mstar/\msun), NIRCam FoV, BaSTI & $6.58\pm0.01$\\
\hline
\hline              
\end{tabular}
\end{center}
\vspace{-10pt}
\tablecomments{The 50\% completeness limits listed for the NIRCam filters are for the full field of view whereas for the ACS filters they are based on the data that overlaps with the NIRCam imaging. See text for details. Parenthetical 50\% completeness limits listed are in absolute magnitudes. The stellar masses reported under the JWST observations are the total mass formed in the full NIRCam field of view and overlapping footprint with ACS, as noted. Similarly, under the HST observations, it is the stellar mass formed in the overlapping footprint with NIRCam. We also report the stellar mass formed based on simultaneously fitting the NIRCam and ACS CMDs, which we list under `joint'. \ttwentyfive, \tfifty, and \tninety\ refer to the lookback times when 25\%, 50\% and 90\% of the stellar mass was formed based on the SFH reconstructed from the full NIRCam data using the PARSEC and BaSTI stellar libraries. References: (1) \citet{McConnachie2012}; (2) \citet{Schlafly2011}; (3) \citet{Albers2019}; (4) \citet{Urbaneja2008}; (5) \citet{Lee2005}; (6) \citet{Leaman2009}}
\end{table}

The first in-depth look at the time-resolved history of a spiral galaxy has come from measurements of the recent and lifetime SFHs of M31, and have included tracking changes in the different structural features of the galaxy and providing a measure of the chemical evolution history of the system \citep{Richardson2008, Lewis2015, Williams2017, Telford2019}. Similar efforts are under way for the lower-mass spiral M33 \citep{Williams2021}. 

Finally, SFHs have been reconstructed for numerous galaxies outside the Local Group (i.e., at slightly farther distances). The photometric depth achieved with HST imaging is more limited, reaching below the horizontal branch (HB), $\sim4-5$ mag below the tip of the red giant branch (TRGB), for a few galaxies \citep[e.g., Leo P, UGC4483;][]{McQuinn2015a, Sacchi2021}, but more typically $1-3$ magnitudes below the TRGB. Given the depth of such data, SFHs can still be characterized but with significantly less temporal resolution at older look-back times (i.e., the oldest-age time bin can be as large as several Gyr). Yet, as these field galaxies are primarily gas-rich, star-forming systems, the SFHs derived from the shallower imaging are useful in that they robustly constrain recent ($\sim1-2$ Gyr) star formation activity with fine temporal resolution \citep[log($\delta$t) $= 0.10$; e.g.,][]{Aloisi1999, McQuinn2010a, McQuinn2015c, McQuinn2015a, Cignoni2018, Annibali2022, Hunter2022}.

To date, almost all SFH determinations from CMD-fitting have been done using data collected with optical filters, with the notable exceptions of the SFH maps performed on near-infrared data from the ESO/VISTA survey of the Magellanic Clouds \citep{Rubele2018, Mazzi2021}. These studies demonstrate that the CMD-fitting method of SFH recovery works as well for near-infrared filters as for optical filters. The comparison between the NIR-derived and the previous optical-derived SFH maps by \citet{Harris2009} showed good quantitative agreement at all ages larger than $\sim10^8$ yr. In some small areas of the Magellanic Clouds, near-infrared photometry has already shown clear advantages over optical photometry by providing sharper CMD features in regions severely affected by differential extinction and by better sampling the evolved stars (especially the age-sensitive sub-giant branch) in regions with a high total extinction. The potential advantages of near-infrared filters have been little explored outside the Magellanic Clouds, mainly because of the intrinsic difficulty to reach deep magnitudes using ground-based near-infrared cameras. This difficulty can now be superseded with the JWST which provides a pathway for SFH work from deep near-infrared CMDs in targeted fields, which we explore in detail. Looking to the future, the large-area near-infrared observations from the Nancy Grace Roman Space Telescope will offer the opportunity to recover SFHs of galaxies residing in many different environments and to study stellar population structures that can span large physical scales, such as stellar streams and galaxy halos.

\subsection{JWST's Advantages for SFHs}
JWST's sensitivity, high resolution, and infrared capabilities enable the robust reconstruction of SFHs of nearby galaxies more efficiently, out to larger distances, and in different galactic environments than is possible with HST. JWST's larger aperture (6.5m vs.\ HST's 2.4m mirror) enables imaging of resolved stars to photometric depths below the oMSTO in galaxies {\em outside} the Local Group. This capability opens up the exploration of the early history of galaxies in varied environments, including isolated environments that provide unique laboratories for probing secular evolution and testing galaxy formation models. JWST's higher resolution and wavelength coverage make morphological and color-based indicators more effective at removing background galaxies, thereby improving the purity of photometric catalogs and SFH fits \citep{Warfield2023}. Cooler stars are intrinsically brighter in the near infrared, which increases observing efficiency when targeting evolved stars (such as RGB and asymptotic giant branch, or AGB, stars) and, importantly for ancient SFHs, oMSTO stars. For example, a metal-poor oMSTO star (\mstar\ $= 0.76$ \msun; 13 Gyr; [M/H] $= -1.4$) has absolute magnitudes of 4.4, 3.6, 3.5, and 3.0 in the F475W, F814W, F090W, and F150W filters, respectively (based on the PARSEC stellar library); this is a gain in brightness of $\sim1.4$ mag moving from the F475W to the F150W filter. Finally, JWST offers potential gains in accurately and precisely recovering SFHs in metal-rich galaxies with significant dust reservoirs that have correspondingly higher extinction.

\begin{figure}[t!]
\begin{center}
\includegraphics[width=0.48\textwidth]{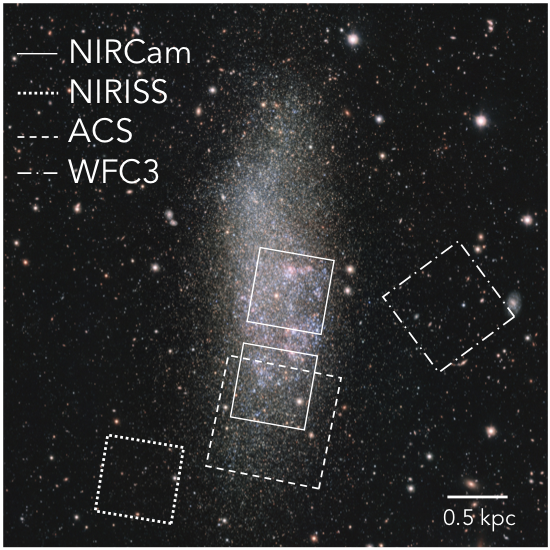}
\end{center}
\caption{Ground-based optical image of WLM (image credit: ESO, VST/Omegacam Local Group Survey) with footprints of the NIRCam (solid), NIRISS (dotted), HST ACS (dashed), and HST WFC3/UVIS (dashed-dotted) fields overlaid.
}
\label{fig:fov}
\end{figure}

\subsection{The SFH Goals of the JWST Early Release Science Resolved Stellar Populations Program}
The JWST Early Release Science Resolved Stellar Populations program (ERS-1334) was designed to provide data reduction and analysis tools for resolved star science with JWST and to test JWST's ability to recover lifetime SFHs from resolved stars, paving the way for future studies. The details on the overall ERS program are presented in \citet{Weisz2023}, with individual science results to be reported separately in upcoming papers from the ERS team. Here, we present the first SFH results from JWST ERS data using the well-established CMD-fitting technique as implemented in {\sc match} \citep[e.g.,][]{Dolphin2002, McQuinn2010a, Weisz2011a, Skillman2017, Savino2023}.

The aims of this SFH analysis are fourfold. First, we will quantify the efficacy of recovering SFHs with the F090W - F150W filter combination. These filters were chosen over the F070W and F115W filters to balance the competing needs of a wide color baseline (which improves metallicity sensitivity in the CMD), signal-to-noise requirements, and the sampling of the PSF. Second, we will test the sensitivity and accuracy of stellar evolution libraries for recovering SFHs and age-metallicty relations or AMRs (i.e., color$-$T$_{\rm eff}$ transformations and photometric calibrations) with the JWST near-infrared data. Third, we will compare the SFH results derived from the individual HST and JWST CMDs with a SFH derived by {\em simultaneously} fitting the JWST and HST CMDs. Finally, we will explore the stellar mass assembly of WLM and compare with other gas-rich low-mass galaxies for which the requisite deep imaging data have been obtained.

\subsection{The star-forming, Local Group galaxy WLM}
The target selected for the JWST Near Infrared Camera (NIRCam) based SFH is the well-studied galaxy Wolf–Lundmark–Melotte (WLM), a gas-rich, low-mass \citep[\mstar\ $=4.3 \times 10^7$ \msun;][]{McConnachie2012} system located near the outskirts of the Local Group \citep[at a distance of $=968^{+41}_{-40}$ kpc;][]{Albers2019}. An overview of WLM's properties is provided in Table~\ref{tab:properties}. 

WLM was chosen primarily because of its proximity, isolation and the rich suite of existing data sets. Specifically, deep HST Advanced Camera for Surveys \citep[ACS;][]{Ford1998} imaging of the stellar disk is available that overlaps with the NIRCam field of view (see Section~\ref{sec:data}). Although the NIRCam data are deeper than the ACS data (50\% completeness limits of M$_{F090W} = +4.30$ mag vs.\ M$_{F475W} = +3.35$), the photometric depth of the ACS data still reaches below the oMSTO. Thus, the SFH derived from the ACS data facilitates a detailed comparison with the SFH derived here from the NIRCam data. 

\begin{figure*}[t!]
\begin{center}
\includegraphics[width=0.48\textwidth]{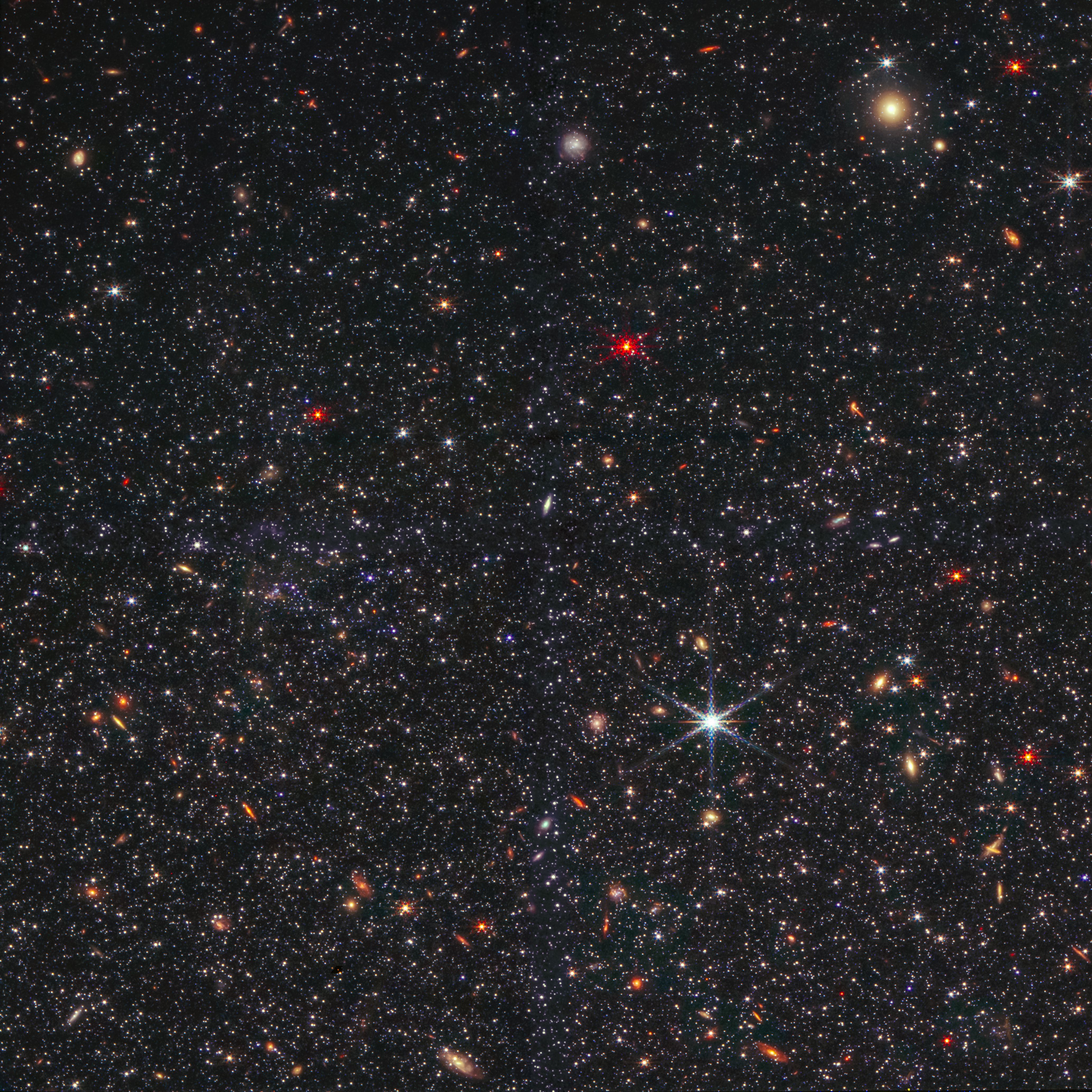}
\includegraphics[width=0.48\textwidth]{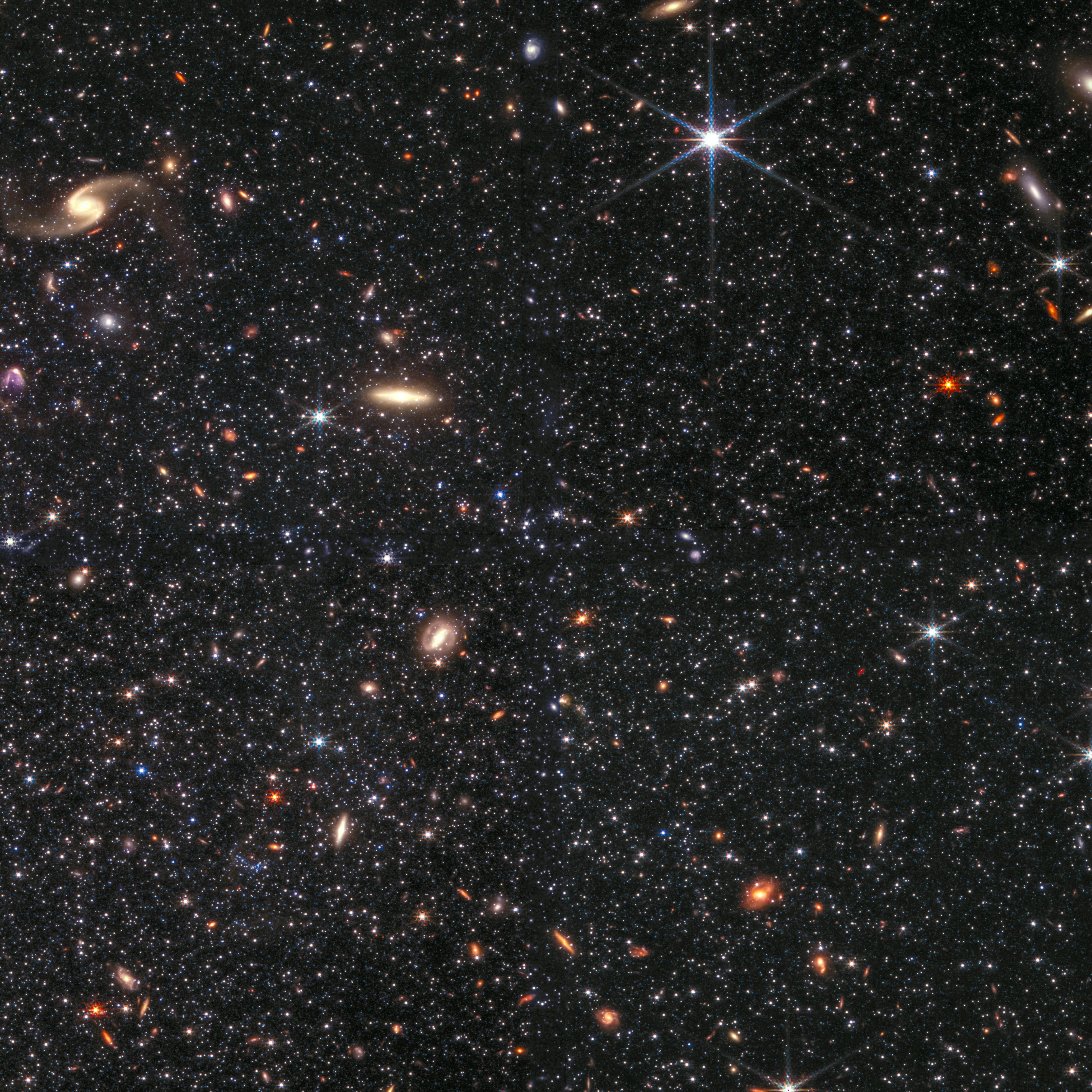}
\end{center}
\caption{JWST NIRCam images from Module A (left) and Module B (right) of the observed field in WLM using F090W (blue), F150W (green), and an average of F250W and F430W (red). Each module is 2.2\arcmin\ on a side, which is equivalent to a physical scale of 0.6 kpc at the distance of WLM. Image Credit: NASA, ESA, CSA, Kristen McQuinn (Rutgers University), Zolt G. Levay, Alyssa Pagan (STScI).}
\label{fig:image}
\end{figure*}

From the previously published ACS-based SFH, it appears that the central part of WLM formed $\sim$50\% of its stellar mass within the last $\sim5$ Gyr, and only $\sim20$\% of its stellar mass at early times \citep{Albers2019}. This implies that WLM experienced a period of quiescence post-reionization ($z\sim6$), which has only been noted in the two other gas-rich, isolated low-mass (\mstar\ $<10^8$ \msun) galaxies in the Local Group with the requisite data quality, Leo~A and Aquarius \citep{Cole2007, Cole2014}. Such a gap in the SFH of low-mass galaxies is somewhat surprising, but given the small sample size, it is unclear whether a pause in star formation is common or atypical. Regardless, the NIRCam data can provide independent confirmation of the HST-based SFH results. Spectroscopic observations of individual stars are also available, which provide robust measurements of the stellar metallicity distribution in WLM \citep{Leaman2009} allowing us to test the AMR derived as part of the CMD-fitting approach. 

This paper is organized as follows. We briefly summarize the observations and data processing of the JWST and the HST data in Section~\ref{sec:data}. The SFHs methodology and results, including analysis of the SFHs derived from stars in the overlapping footprint between NIRCam and ACS and a comparison of the AMR with an independently measured MDF, are presented in Sections~\ref{sec:sfhs} and \ref{sec:results}. We summarize our findings in Section~\ref{sec:conclusions}.

\section{Observations and Data Reduction}\label{sec:data}
\subsection{Overview of JWST NIRCam Data}
Shown in Figure~\ref{fig:fov}, JWST observed a central region of the stellar disk in WLM with NIRCam \citep{Rieke2023} as the primary instrument (solid lines) and an outer field with Near Infrared Imager and Slitless Spectrograph \citep[NIRISS;][]{Doyon2023} in parallel (dotted lines). For NIRCam, we selected orient constraints such that one NIRCam module overlapped a region with the deepest optical HST imaging available \citep[dashed lines, HST-GO-13678;][]{Albers2019}, while the other module overlapped a region with HST UV, optical, and IR imaging from various programs (not shown; PIDs: HST-GO-6798, HST-GO-11079, HST-GO-15275, HST-GO-16162), and ALMA observations where CO clouds were identified \citep{Rubio2015}. This carefully chosen areal coverage with such comprehensive, multi-wavelength ancillary data enables multiple science goals including: a comparison of the SFHs derived from JWST and HST presented in this work, a study of spatial variations in the SFHs, an investigation of the evolved stars, and an exploration of the stellar populations co-spatial with the CO clouds. Dictated by the orient constraints used for the NIRCam placement, the NIRISS field covered an outer region of the galaxy to the southeast. Here, we focus on the NIRCam data; analysis of the less-populated NIRISS imaging will be presented in a future paper. 

Table~\ref{tab:properties} lists the coordinates of our observed fields, and the total integration times per filter for NIRCam from ERS-1334. These data were obtained from the Mikulski Archive for Space Telescopes (MAST) at the Space Telescope Science Institute. The specific observations analyzed can be accessed via \dataset[DOI: 10.17909/t6pm-x063]{https://doi.org/10.17909/t6pm-x063}. We refer the reader to \citet{Weisz2023} for a detailed description of the observing strategy. Note that while the full NIRCam data includes imaging from the Long Wavelength (LW) F250W and F430M filters, we only use the Short Wavelength (SW) F090W and F150W filters to derive the SFH as the data are deeper and this filter pair is more sensitive to changes in stellar ages.

Figure~\ref{fig:image} presents color images for the two NIRCam modules created using the F090W filter (blue), F150W filter (green), and the sum of the F250M and F430M filters (red). The image demonstrates the exquisite resolution and depth of the NIRCam imaging that will be exceptional for resolved star science in nearby galaxies. Careful inspection of the fields reveal changes in the stellar populations, from small variations in the stellar density, to stellar sources still somewhat embedded in their nascent clouds, to extremely red stars that are dusty, AGB candidates.

\subsection{NIRCam Photometry and Artificial Star Tests}
We performed point-spread function (PSF) photometry on the NIRCam images using the photometry software package {\tt DOLPHOT}\footnote{http://americano.dolphinsim.com/dolphot/nircam.html} \citep{Dolphin2000, Dolphin2016} with the new beta-version of a JWST NIRCam-specific module (Weisz et al., submitted). As inputs, we used the F150W {\tt stage 3} drizzled {\tt i2d.fits} file from the JWST data pipeline as the astrometric reference image for identifying point sources, and the {\tt cal.fits} files processed with the JWST calibration pipeline version 1.9.6 and the {\tt pmap} 1075 reference files to measure the flux of each identified source. 
The {\tt cal.fits} files and the {\tt i2d.fits} file were internally aligned by the JWST pipeline.

Our photometric reduction with {\tt DOLPHOT} used model PSFs generated with the WebbPSF simulation tool (version 1.1.0). The WebbPSF tool adopts in-flight optical performance data and incorporates time-dependent optical path delay (OPD) maps based on the telescope configuration. We used the OPD map of the NIRCam instrument around the time our observations were taken. Despite being well-matched to the telescope conditions, the WebbPSF models appear to be slightly more sharply peaked than the actual PSF of the images (i.e., the modelled PSFs have more flux in the central regions) at the $\sim2$\% percent level, which currently limit the relative precision of our photometry (see Weisz et al., submitted for details).

Before running photometry, we masked out bad pixels, identified saturated pixels, and applied pixel area maps using the routine {\tt nircammask} included as part of the {\tt DOLPHOT} distribution. The photometric parameters used to run {\tt DOLPHOT} were chosen after extensive testing of the photometric recovery from NIRCam (Weisz et al., submitted). The parameters are very similar to those adopted for HST WFC3/IR imaging in the PHAT program \citep{Williams2014}. These {\tt DOLPHOT} parameters were optimized for the NIRCam imaging based on the current PSF models and may need adjusting if the PSF model changes. 

The photometry was filtered for well-recovered point sources. We required a signal-to-noise ratio $\geq$4 in each filter, {\tt sharp}$^2$ $\leq$ 0.0225 in each filter, and {\tt crowd} $\leq$ 0.5 in each filter. The sharpness parameter is a measure of how peaked or broad a source is relative to the PSF and helps to reject image artifacts and background galaxies, respectively. The crowding parameter measures how much brighter a source would have been if stars nearby on the sky had not been fit simultaneously. 

\begin{figure}[t!]
\begin{center}
\includegraphics[width=0.46\textwidth]{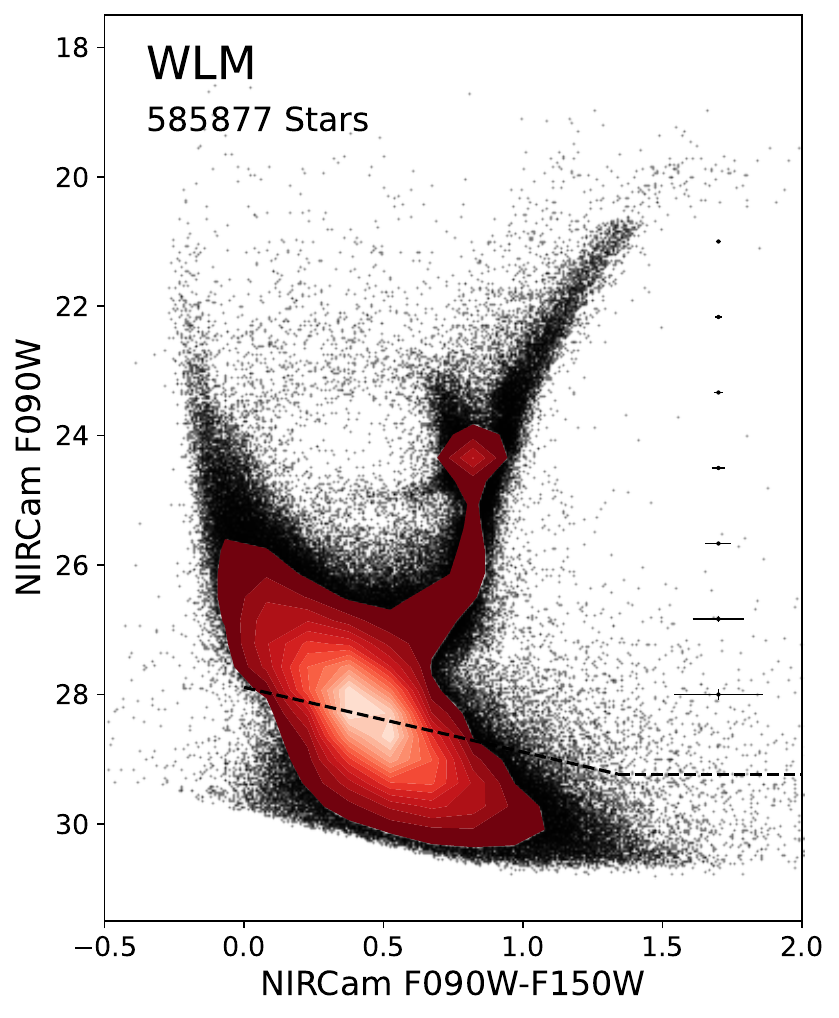}
\end{center}
\caption{CMD from the NIRCam stellar catalog recovered from both modules which includes close to 600,000 sources after applying our quality cuts. Representative uncertainties per magnitude are shown on the right. The red contours are a 2-D histogram of CMD regions with higher source densities; lighter shades of red trace the peak concentration of sources. The CMD is well-populated by stars in all major phases of evolution, providing an excellent data set to test the SFH recovery from NIRCam data. The CMD is the deepest obtained to date for any galaxy that is not in close proximity to the Milky Way. The 50\% completeness limit is marked with a dashed line; see Table~\ref{tab:properties} for values.}
\label{fig:jwst_cmd}
\end{figure}

\begin{figure*}[t!]
\begin{center}
\includegraphics[width=0.95\textwidth]{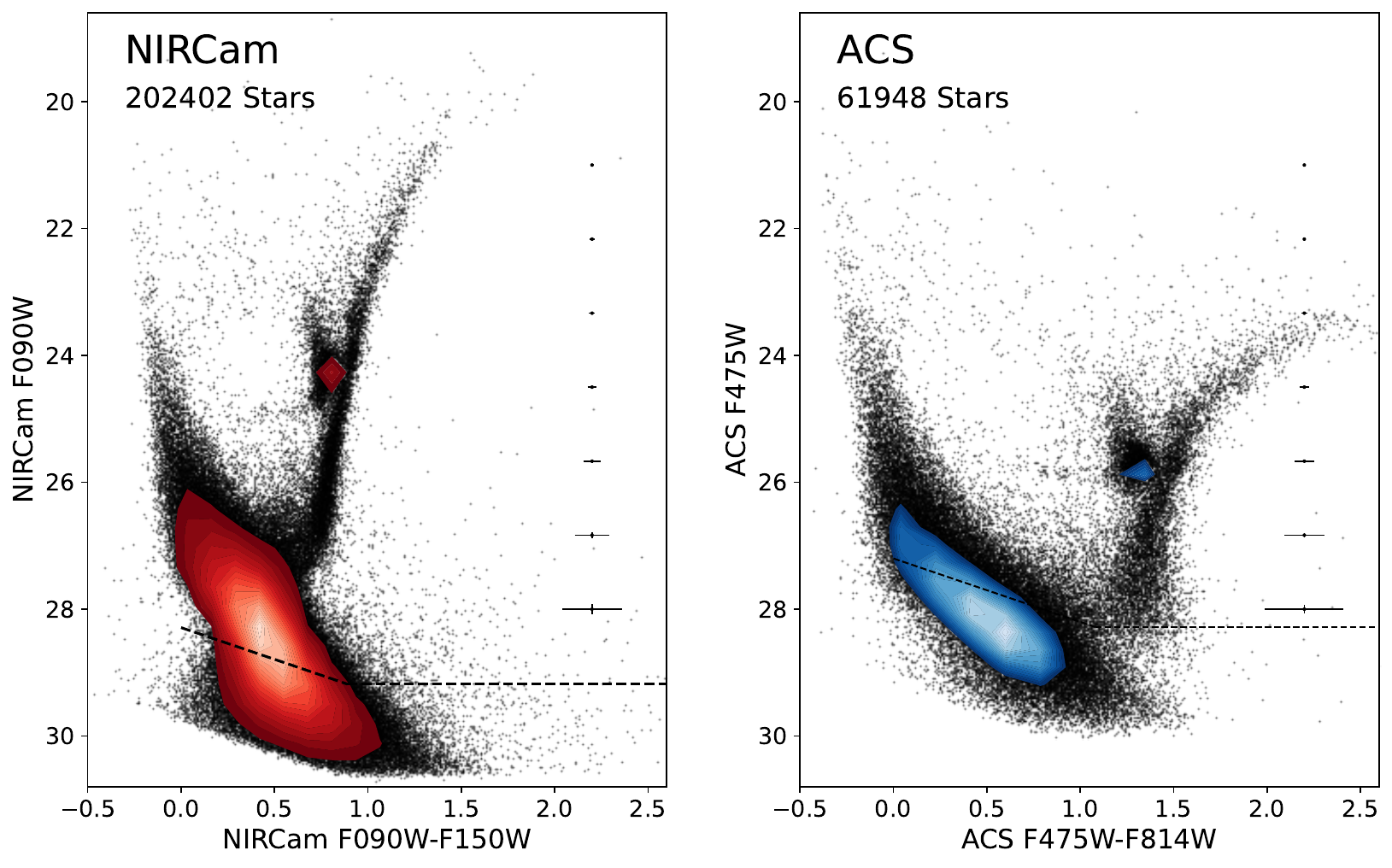}
\end{center}
\caption{CMDs of stars within the NIRCam - ACS matched field of view. The NIRCam CMD (left) is $\sim$2 mag deeper than the ACS CMD (right) and has a factor of $\sim3$ more stars; 50\% completeness limits are marked as dashed lines on the respective CMD. Note that the completeness in the NIRCam CMD is slighter deeper than the data from the full field of view shown in Figure~\ref{fig:jwst_cmd} due to the change in object density between Modules A and B. Similar to Figure~\ref{fig:jwst_cmd}, the contours are a 2-D histogram of CMD regions with higher source densities; lighter shades of color trace the peak concentration of sources. While the color baseline of the F090W$-$F150W and F475W$-$F814W filter combinations are similar, the features in the NIRCam CMD become narrower as the near infrared wavelengths approach the Rayleigh-Jeans regime. However, even with the narrower features in the F090W and F150W filters, the different stellar evolution branches are well-separated in the NIRCam CMD.}
\label{fig:cmds_jwst_hst}
\end{figure*}

We also ran a large number (2.3M) of artificial star tests (ASTs) per NIRCam module to measure the bias and completeness of the data and the photometric errors. While fewer ASTs (of order a few 100k) would have been sufficient for our SFH-recovery purposes, we ran such a large number of ASTs for data exploration and testing purposes for this first SFH fit with NIRCam data. Artificial stars were injected throughout the images with a flat spatial distribution (similar to the distribution of the stellar populations in each NIRCam module) and then recovered with {\tt DOLPHOT} using the same parameters set for the photometry. We then applied the same quality cuts used for filtering the photometry and marked any star that did not meet our quality requirements as unrecovered. Based on the ASTs, the 50\% completeness limits averaged over the full NIRCam stellar catalogs are $m_{F090W}=29.23$ (M$_{F090W} = +4.30$) mag and $m_{F150W}=27.89$ (M$_{F150W} = +2.96$) mag; these values are also listed in Table~\ref{tab:properties}.

Figure~\ref{fig:jwst_cmd} presents the NIRCam CMD with close to 600,000 well-recovered point sources; mean $1\,\sigma$ uncertainties per magnitude bin are shown on the right. Stellar populations with a wide range of different ages and masses are well-represented in the WLM field, creating a comprehensive CMD with various branches of stellar evolution that can be compared with existing data and with different stellar models. Specifically, the CMD includes younger stars in the upper main sequence and blue/red helium burning sequences, intermediate and older age stars on the red giant and asymptotic giant branches and in the red clump, and older stars in the horizontal branch, sub-giant branch, and lower main sequence. 

\subsection{Co-Spatial HST ACS stellar catalogs} 
The JWST NIRCam pointing intentionally targeted a field in WLM that overlapped with existing deep HST ACS optical imaging \dataset[DOI: 10.17909/gs15-rb46]{https://doi.org/10.17909/gs15-rb46}. The region in common between the two data sets, seen in Figure~\ref{fig:fov}, allows for a well-matched and detailed comparison of the resolved star photometry recovered from each telescope and the SFHs reconstructed separately from infrared vs.\ optical data. In addition, the overlap in areal coverage also allowed us to perform a joint photometric reduction of the JWST and HST data together and recover the SFH by {\em simultaneously} fitting the infrared and optical CMDs. Here, we briefly describe the reduction of the ACS data and the creation of stellar catalogs from the NIRCam - ACS overlapping field of view; we describe the simultaneous photometric reduction in  Section~\ref{sec:joint_phot}.

The HST ACS observations were taken as part of HST-GO-13768, described in detail in \citet{Albers2019}, and include imaging in the F475W and F814W filters. The data were processed with the same tools and methodology used on the JWST data. Specifically, the HST-only photometry was performed on the HST pipeline-processed {\tt flc.fits} files using the ACS module of {\tt DOLPHOT} with a F814W drizzled image. The same quality cuts used for the NIRCam data were applied to create the final ACS stellar catalog. These cuts are similar to those used on the ACS data in \citet{Albers2019}, except we use a slightly more stringent cut on sharpness (which increases the purity of the catalog at the expense of completeness). Approximately 500,000 ASTs were run on the ACS data in the region overlapping with the NIRCam data. 

From the full stellar catalogs of the NIRCam Module B (the only module that overlapped with ACS) and the ACS data, we selected sources that lie within the common footprint, excluding sources that fall in any of the chip gaps. Figure~\ref{fig:cmds_jwst_hst} presents the CMDs of these stars, with representative uncertainties per magnitude bin. The photometry from both NIRCam and ACS reach below the oMSTO, although, as expected given the differences in the telescopes and observing strategies, the NIRCam data has a higher S/N at faint magnitudes. Quantitatively, the 50\% completeness limits for the NIRCam data are $m_{F090W}=29.18$ (M$_{F090W} = +4.25$) mag and $m_{F150W} = 28.29$ (M$_{F150W} = +3.36$) mag, compared with $m_{F475W} = 28.28$ (M$_{F475W} = +3.35$) mag and $m_{F814W} = 27.20$ (M$_{F814W} = +2.27$) mag for the ACS data; the ACS values are also listed in Table~\ref{tab:properties}. The completeness limit for the F150W filter in this region of WLM is somewhat fainter than that in the full NIRCam field of view, which reflects the overall lower {\em object} crowding in this region (Weisz et al., submitted). This difference in crowding has a greater impact on the F150W photometry due to the wavelength coverage and the larger PSF in this bandpass relative to the F090W.  

The total number of high-confidence stars recovered photometrically in the NIRCam data is $\sim3$ times higher than the number in the ACS data in the matched field of view. The higher star counts reflect the higher completeness of the NIRCam data at faint magnitudes where star counts naturally increase for any typical initial mass function (IMF). It also reflects that the lower main sequence stars are brighter in the near infrared than in the blue F475W filter. For example, an F150W absolute magnitude of $+4.3$ corresponds to an oMSTO star with a mass of 0.65 \msun\ (based on the PARSEC library) whereas in the F475W filter this same absolute magnitude corresponds to a star with 0.72 \msun. At brighter magnitudes, where neither data set is significantly impacted by incompleteness (e.g., F090W $<$ 26 mag), the number of recovered stars from the two instruments are approximately equal. We compare the SFHs derived separately from these catalogs in Section~\ref{sec:sfh_jwst_hst}. 

Besides photometric depth, completeness, and star counts, there are other differences in the CMDs. Most notable are the slightly different CMD architectures due to differences in the narrower CMD features in the near infrared. While the width of the color baselines in the F090W$-$F150W and F475W$-$F814W filter combinations are similar, the features in the NIRCam CMD become narrower as the near infrared wavelengths approach the Rayleigh-Jeans regime of a star's spectral energy distribution for bluer sources.

\subsection{Simultaneous Photometric Reduction of the NIRCam and ACS Images}\label{sec:joint_phot}
We also performed simultaneous, cross-observatory PSF photometry using {\tt DOLPHOT} on the NIRCam and ACS data. We used the same methodology and parameters described above, with the F090W {\tt i2d.fits} file as the reference frame, the {\tt cal.fits} files as the science images for NIRCam, the {\tt flc.fits} files for ACS, and applied the same quality cuts to the output catalogs. For the ASTs, we generated an input list of $\sim1$M sources with realistic spectral energy distributions across the wavelength range covered by the F475W - F150W filters. We applied the same treatment to the ASTs as before. The stellar catalogs are similar to those generated from the individual photometry runs, but with slightly fewer stars passing the quality cuts. We recovered 177,000 stars in the NIRCam filters, compared with 202,000 in the NIRCam-only catalog, and 58,000 in the ACS filters, compared with 62,000 in the ACS-only catalog. The 50\% completeness limits for the multi-observatory photometry are $m_{F090W}=28.77$ mag, $m_{F150W} = 28.09$ mag, $m_{F475W} = 27.74$ mag, and $m_{F814W} = 26.86$ mag. These are somewhat brighter than the limits from photometry performed on each data set separately, similar to what has been noted when processing, for example, HST ACS and WFC3/IR data simultaneously \citep{Williams2014}.

\begin{figure}[t!]
\begin{center}
\includegraphics[width=0.46\textwidth]{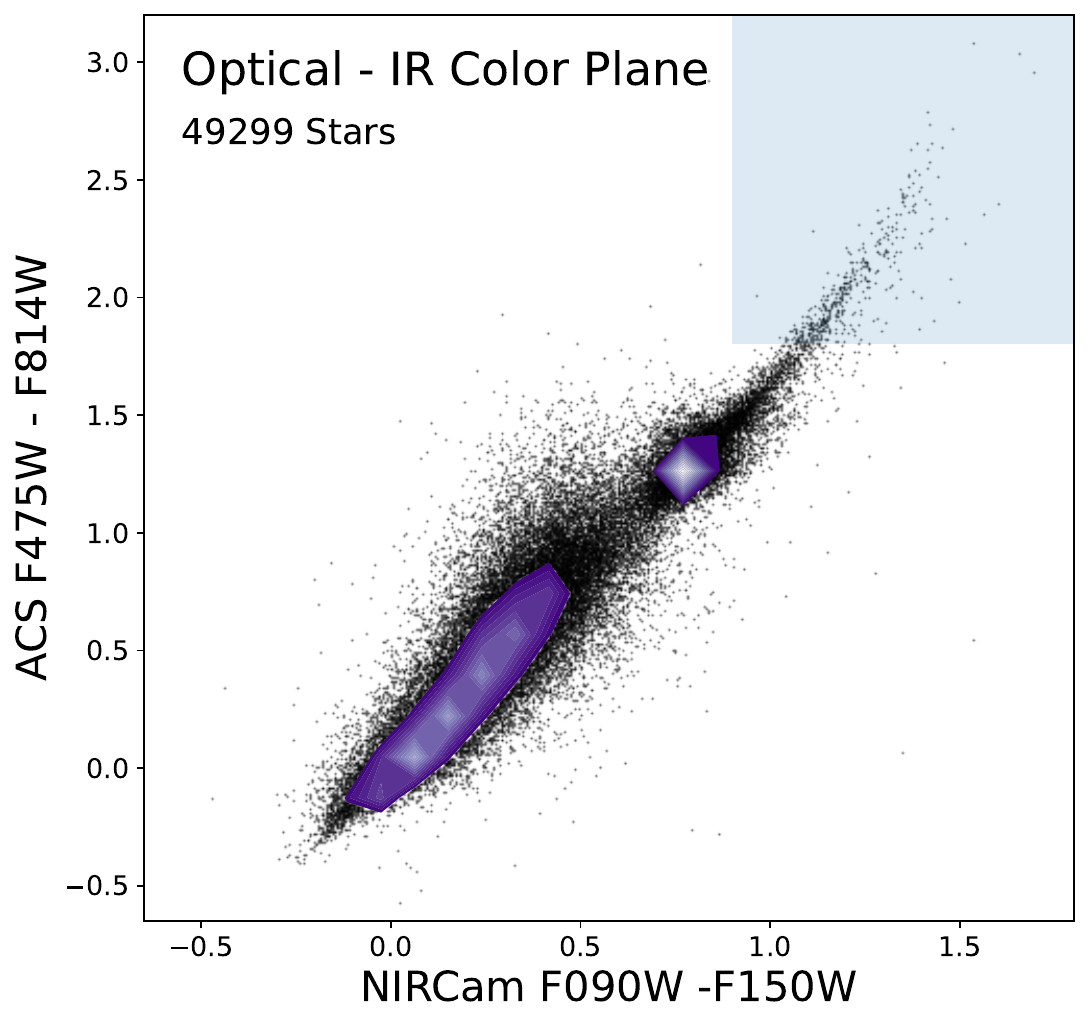}
\end{center}
\caption{The NIRCam IR - ACS optical olor-color plane populated with point sources recovered in both NIRCam filters (F090W, F150W) and both ACS filters (F475W, F814W) from the cross-observatory photometric reduction. Dominant features trace the main sequence, the red clump, and the RGB (see text for details); the increase in photometric uncertainties for fainter sources in the middle color-color range is also seen. The blue shaded region is where the majority of foreground stars from the MW are expected to lie; as expected given the high Galactic latitude, there is no obvious population of foreground contaminants in the cleaned catalogs.}
\label{fig:color_color}
\end{figure}

Figure~\ref{fig:color_color} presents the NIRCam IR - ACS optical color-color plane with point sources that are well-recovered in all four filters from the simultaneous photometric reduction. The tightness of the stellar sequences in the color-color emphasizes the high-fidelity of the photometry. Evident in the plot are a number of features including: (i) the red clump over-density at F090W-F150W $\sim0.8$, F475W-F814W $\sim1.3$; (ii), the tight RGB sequence at F090W-F150W $\>0.8$, F475W-F814W $\>1.3$; (iii) the richly populated main sequence at F090W-F150W $\ltsimeq0.4$, F475W-F814W $\ltsimeq0.6$; and (iv) the photometric spread at fainter magnitudes at F090W-F150W $\sim0.3- 0.6$, F475W-F814W $\sim0.4-1.5$. We do not see a prominent galaxy sequence in the color-color plot, which demonstrates the power of the high-resolution of NIRCam and ACS imaging combined with accurate PSF photometry with {\tt DOLPHOT} and carefully selected quality cuts. The shaded blue at redder optical and IR colors represents the region where, if present, the majority of foreground MW stars would lie. Given the higher Galactic latitude of WLM (b$= -73^{\circ}$), we do not expect a significant population of foreground stars, which is confirmed by the low source counts in this region that are not coincident with the RGB sequence or the few redder sources that are likely AGB stars. Note that bluer, foreground dwarfs are predominantly brighter than the magnitude ranges in our ACS and NIRCam CMDs and, therefore, not a concern.

\section{The Star Formation History Methodology}\label{sec:sfhs}
\subsection{CMD-fitting Technique}\label{sec:sfh_method}
The SFHs and AMRs were derived using the numerical CMD-fitting software \textsc{match} \citep{Dolphin2002}. \textsc{match} employs the well-established technique of comparing an observed CMD to a grid of synthetic simple stellar populations (SSPs) of different ages and metallicities \citep[e.g.,][]{Tolstoy2009}. The SSPs generated assume an IMF, a binary fraction, and a set of stellar evolution isochrones, and take into account photometric completeness determined from the ASTs. The modelled CMDs are linearly combined until the best-fit to the observed CMD is found based on a Poisson likelihood statistic.

The SFHs were derived using two stellar evolution libraries with solar-scaled abundance patterns updated with NIRCam in-flight filter transmission curves and Vega zeropoints, namely PARSEC \citep{Bressan2012} and BaSTI \citep{Hidalgo2018}. We assumed a Kroupa IMF \citep{Kroupa2001} and a binary fraction of 35\% with a flat mass ratio distribution for the secondaries. The SFH solutions were fit with an age grid of log(t)$ = 6.6-10.15$ using a time resolution log($\delta$t$) = 0.1$ dex for ages less than log(t)$ = 9$ and log($\delta$t$) = 0.05$ dex for older ages, and a metallicity grid [M/H]$= -2.2$ to $-0.3$ with a resolution of 0.1 dex. The upper bound on the metallicity range was chosen to be well-above the spectroscopically measured present-day metallicity of [Fe/H] $\sim-0.8$ \citep{Bresolin2006, Urbaneja2008}. We also enforced a restriction that the metallicity monotonically increases in time. For the separate SFH fits to the NIRCam data that overlap with the ACS field,  we also required that the AMR from NIRCam matched the best-fitting AMR from the ACS data, thereby providing a one-to-one comparison of the recovered SFHs. We also experimented with fitting the NIRCam data without requiring a matching AMR to the ACS data. The resulting NIRCam SFHs were in excellent agreement with the results requiring a matched AMR. The independently fit AMRs were also in good agreement with the ACS AMRs with the initial and final values of [M/H] and overall shape of the AMR a close match for each stellar library. 

We limited the fitting to stars brighter than the 50\% completeness limits in each filter in a  given CMD, as listed in Table~\ref{tab:properties}. We note that models of the AGB phase of stellar evolution are uncertain, but given the small number of AGB stars in the data, the inclusion of these stars in the catalogs does not impact the derived SFHs. Following the same procedure used in \citet{Albers2019}, we adopted the distance modulus of 24.93 mag measured from the TRGB luminosity in the HST ACS data that overlap with our NIRCam field. We also assumed the same foreground extinction value of $A_V = $0.10 mag based on Galactic dust maps from \citet{Schlegel1998} with re-calibration by \citet{Schlafly2011}, an $R_V = 3.1$, and the reddening law of \citet{Fitzpatrick1999}. WLM is at a high Galactic latitude where foreground contamination is expected to be minimal, as demonstrated in Figure~\ref{fig:color_color}. Nonetheless, following best practices, we included a model of foreground stars in the Milky Way in the fits, generated with the TRILEGAL simulation tool \citep{Girardi2012}. Confirming that foreground stars are minimally present in our stellar catalogs, the SFH solutions with and without the inclusion of the foreground model were identical.

\begin{figure*}
\gridline{\fig{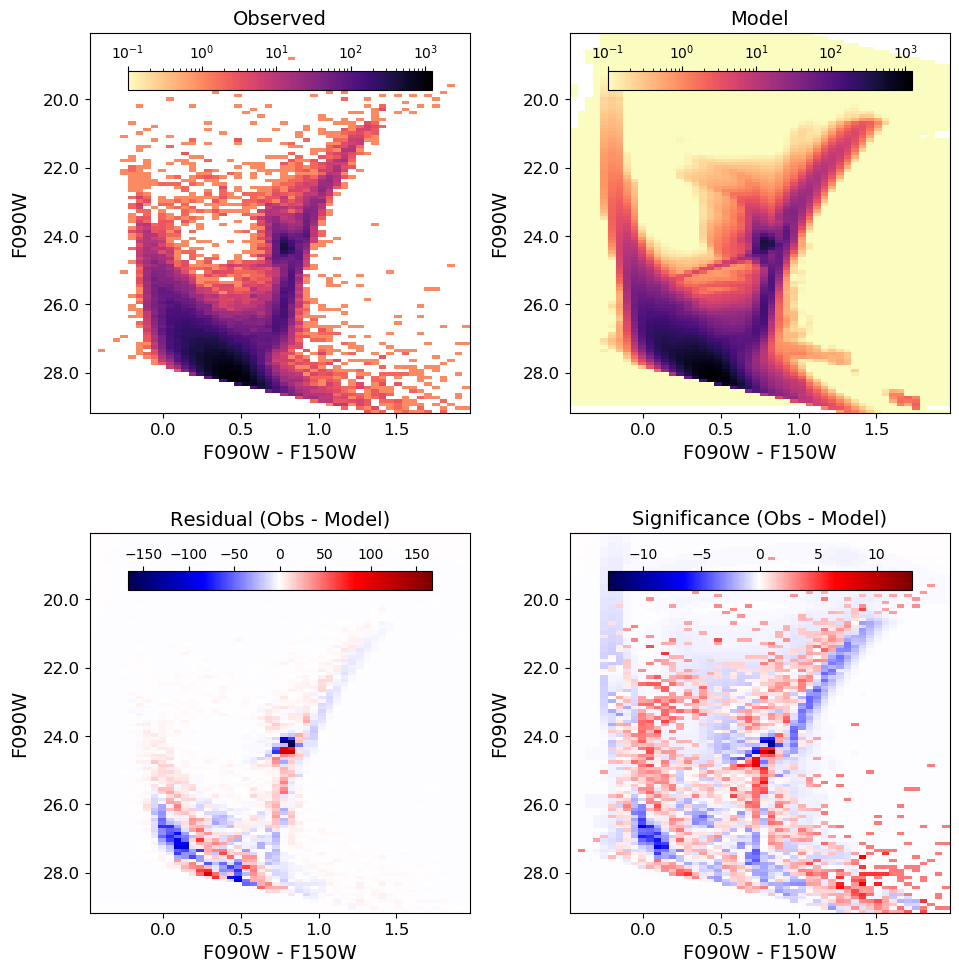}{0.45\textwidth}{(a) JWST NIRCam data fit with the PARSEC library}
    \fig{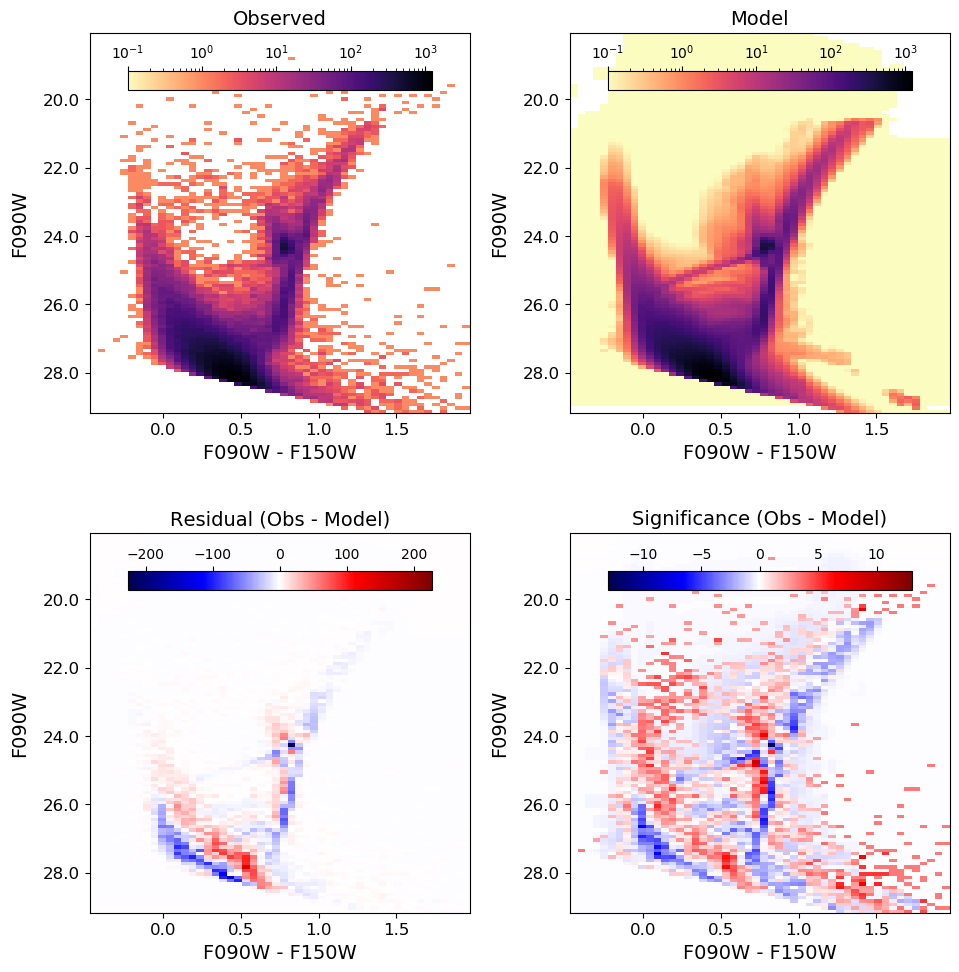}{0.45\textwidth}{b) JWST NIRCam data fit with the BaSTI library}}
\gridline{\fig{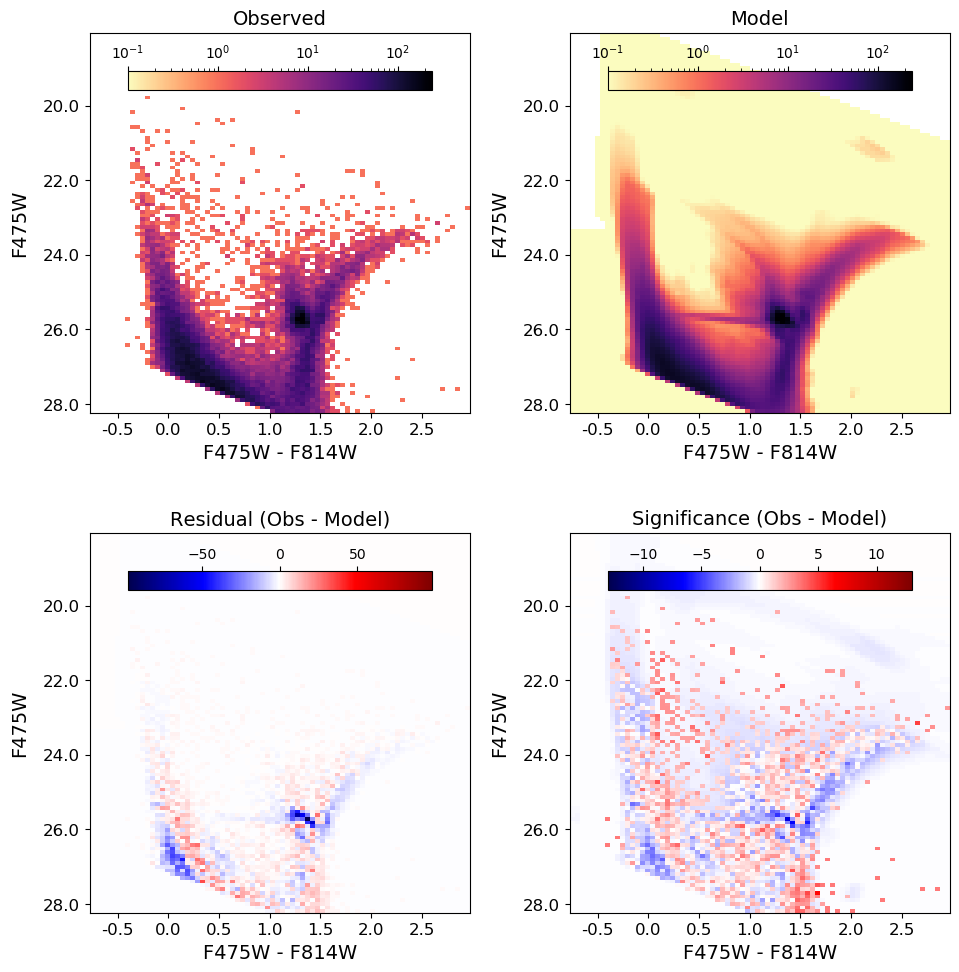}{0.45\textwidth}{(c) HST ACS data fit with the PARSEC library}
     \fig{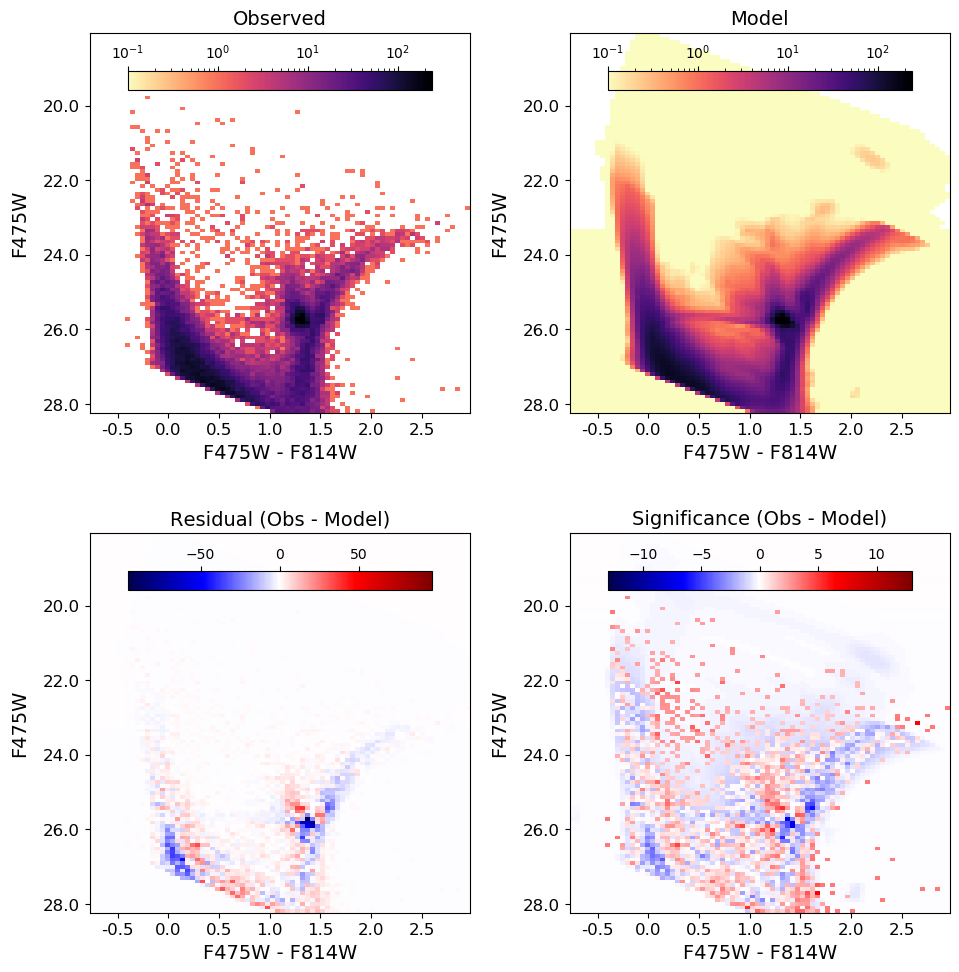}{0.45\textwidth}{(d) HST ACS data fit with the BaSTI library}}
\caption{Example fits to the NIRCam (subplots a, b) and ACS (subplots c, d) stellar catalogs in the common field of view. For each data set, fits are shown using PARSEC (subplots a, c) and BaSTI (subplots b, d) stellar libraries. For each data set $+$ stellar library combination, we show four sub-panels. The top two panels show the observed data in Hess diagram format and the modelled Hess diagram. The bottom two panels show the residual (the observed Hess diagram$-$model Hess diagram) and residual significance (defined as the observed Hess diagram$-$model Hess diagram weighted by the variance in each pixel). For direct comparison between fits, the residual significance plots are fixed to the same colorbar range.}
\label{fig:residuals}
\end{figure*}

Statistical uncertainties due to the finite number of stars in the CMD were estimated using a Hybrid Markov Chain Monte Carlo approach \citep{Dolphin2013}. The SFH solutions derived with the two stellar libraries show the impact of different assumptions used in the stellar models, and therefore give some indication of the systematic uncertainties due to stellar evolution. However, with only two libraries, these differences likely underestimate the true systematic uncertainties of the solutions. Future work will include fits from additional stellar libraries once they have been updated with the in-flight performance of NIRCam, as well as systematic uncertainty estimates using Monte Carlo simulations that are tuned based on solutions from multiple stellar libraries \citep[e.g.,][]{Dolphin2012}. As a consistency check for our ACS SFH solution, we compared our results to those derived from the full ACS field of view presented in the Appendix of \citet{Albers2019}, which includes fits to the BaSTI and PARSEC stellar libraries used in the present work, as well as the older Padua models \citep{Girardi2010} and the MIST models \citep{Choi2016}. We found the solutions to be in excellent agreement and well within the uncertainties in \citet{Albers2019}.

\subsection{The Best-Fits to the Independent NIRCam and ACS CMDs}\label{sec:residuals}
Figure~\ref{fig:residuals} presents the CMD-fits to the stellar catalogs from the overlapping region in the NIRCam and ACS data. These include independent solutions from the NIRCam and ACS catalogs fit with the PARSEC and BaSTI stellar libraries. Each data set$+$library solution has four subplots, as described in the figure caption. For interpreting the quality of the fits, the most useful diagnostic is the residual significance shown in the lower right for each solution. Note that the scale of the significance color bar has been fixed to the same range so differences between each data$+$model can be directly compared. While we only show the results for the data in the overlapping region between NIRCam and ACS, the general patterns in these residual significance plots are representative of the solutions from the full NIRCam catalogs.

Overall, the modelled CMDs are a very good representation of the data, with the exception of a few regions. Specifically, the red clump (RC) area is not well-modelled, which has been noted in numerous previous studies with ACS optical observations \citep[e.g.,][]{Gallart2005}. It is this area that sets the extrema for the ranges in the residual significance. In addition, the modelled blue helium burning (BHeB) sequences are somewhat redder (i.e., cooler) than the observed sequences. This has also been noted previously based on optical observations \citep[e.g.,][]{McQuinn2011}, and the possible causes of the disagreement have been explored theoretically \citep[e.g.,][]{Tang2014}.

Smaller differences are seen between the models and the data for RGB stars. The PARSEC and BaSTI models create a slightly redder (i.e., cooler) RGB relative to both the NIRCam and the ACS data. A similar offset  was reported at the level of 0.05 mag based on the JWST ERS data on M92 \citep{Weisz2023}; it has also been previously noted for a number of stellar models in near infrared 2MASS broadband filters across a range of metallicities in Galactic globular clusters \citep{Cohen2015}. The slight mismatches between model and data have similar trends in the JWST and HST bands, suggesting there are not inherent biases in using the JWST over HST bands for SFH fitting, although the mismatches are somewhat more pronounced in the JWST residuals plots.

A color offset between the model and data can indicate the presence of internal extinction that should be considered in the fits. As internal extinction can be treated as a free parameter  in {\sc match}, we explored whether adding extinction to the fits would improve the solutions. Briefly, we varied extinction up to 0.50 mag, in increments of $A_V = 0.05$ mag, and iteratively re-fit the CMD with each stellar library. For the BaSTI models, extinction did not reduce the residual significance in the fits. For the PARSEC models, internal extinction of $\sim$0.30 ($\sim$0.15) mag compensated for the offset in color between the modelled and observed RGB sequence for the NIRCam (ACS) data, with slightly lower values in the residual significance. However, while the SFH solutions with and without the extinction were consistent with each other, the inclusion of internal extinction resulted in lower stellar metallicities in the AMRs; these lower metallicities are a poor match to spectroscopic abundances (see below), are inconsistent with the expectation of minimal extinction in such a metal-poor galaxy \citep[e.g.,][]{McQuinn2015c}, and are inconsistent with the lack of internal extinction found based on the ACS data reported in \citet{Albers2019}. Thus, for our final fits, while we continued to include a foreground Galactic extinction of $A_V=0.10$ mag (see \S\ref{sec:sfh_method}), we did not adopt any internal extinction.

\subsection{The Best-Fits to the Simultaneous NIRCam and ACS CMDs}\label{sec:residuals_simultaneous}
We also used the photometry and ASTs run simultaneously on the JWST and HST images to fit the NIRCam and ACS CMDs together. The modelled CMDs and residuals between the observed and modelled data from both libaries were qualitatively similar to the results from fitting the NIRCam and ACS data separately. We show these results in the Appendix, which can be compared to Figure~\ref{fig:residuals}. Qunatitatively evaluating the fits is not straightforward because, on the one hand, using two CMDs in the fit increases the information available to derive a SFH, but on the other hand, the fits are based on different numbers of stars and the more complex data can disallow some solutions. One basis for evaluation is the Poisson likelihood statistic returned by {\sc match}. Using this as a measure of the fit quality, we find the likelihood statistic from modelling the 2-CMDs simultaneously was comparable to the combined values from modelling each CMD independently using the PARSEC library, but slightly worse based on BaSTI library. This would suggest that there are not obvious advantages to using both NIRCam and ACS CMDs to recover a SFH over just a single CMD. Another basis for evaluating the fits is to compare the best-fitting SFHs and AMRs from each method, which are presented in Section~\ref{sec:sfh_jwst_hst}.
\vspace{10pt}

\section{The Star Formation History of WLM}\label{sec:results}
\subsection{SFH and AMR from the NIRCam data}\label{sec:sfh_nircam}
\begin{figure}
\begin{center}
\includegraphics[width=0.45\textwidth]{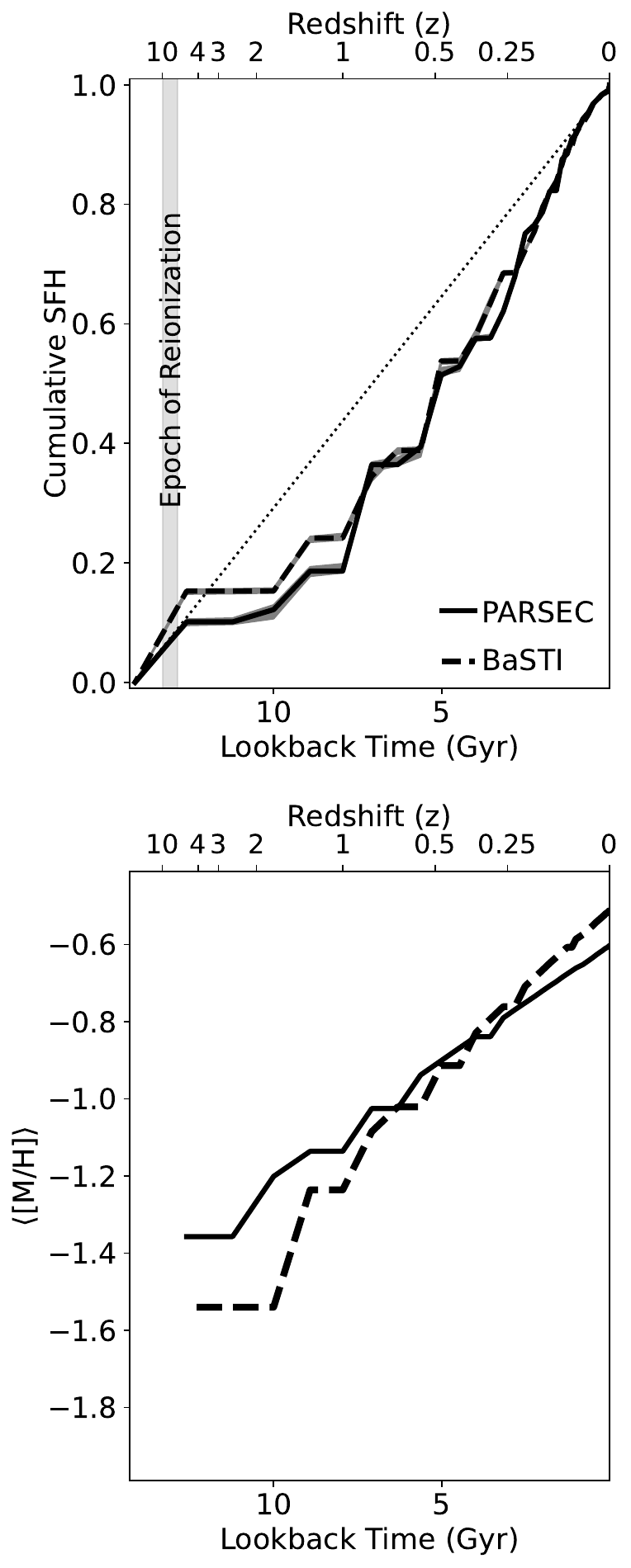}
\end{center}
\caption{Global cumulative SFH (top) and AMR (bottom) derived from NIRCam imaging of the main stellar disk in WLM using the PARSEC (solid lines) and BaSTI (dashed lines) libraries. Shaded regions represent the statistical uncertainties on the fit based on the finite numbers of stars in the CMD. The dotted black line in the top panel represents a constant star formation rate. The earliest value shown for the AMR corresponds to the first time bin with a stellar mass measurement.}
\label{fig:sfh_nircam}
\end{figure}

Figure~\ref{fig:sfh_nircam} presents the best-fitting cumulative SFHs (left) and AMRs (right) from the NIRCam full field of view with solutions using the PARSEC (solid lines) and BaSTI (dashed lines) stellar libraries. The top x-axis converts lookback time to redshift based on the \citet{Planck2020} cosmology. Shading on the SFHs represent statistical uncertainties only. The vertical gray region encompasses the approximate range of the epoch of reionization from $z\sim10-6$, corresponding to a lookback time from 13.3$-$12.9 Gyr. 

The SFH and AMR results from both libraries are in excellent agreement with one another. The differences between SFH solutions give some indication of the systematic uncertainties, but additional stellar libraries with updated JWST zeropoints are needed before the full range of systematic uncertainties can be estimated. We also find that the total stellar mass formed based on the SFH fits from each stellar library are in excellent agreement which provides an additional quantitative measure of the SFH results. These values are listed in Table~\ref{tab:properties}. 

The SFH of WLM (in the region imaged by NIRCam) shows that star formation occurred at early times followed by an interruption post-reionization ($z\sim6$; lookback time of 12.9 Gyr). This pause in star formation extended over $\sim3$ Gyr after which there was a marked increase in the star formation activity. To provide a quantitative measure that can be compared with other galaxies, we calculate the timescales ``$\tau$'' over which different fractions of the stellar mass were formed. Focusing on just the PARSEC models for simplicity and interpolating the SFH in Figure~\ref{fig:sfh_nircam}, we find that $\sim$10\% of the stellar mass formed within a lookback time \tten\ $= 12.61\pm0.06$ Gyr, followed by a pause until $\sim$10 Gyr ago, and then the mass more than doubles reaching \ttwentyfive\ at a lookback time $= 7.63^{+0.05}_{-0.03}$ Gyr; the uncertainties represent statistical uncertainties determined using a Markov Chain Monte Carlo approach. The stellar mass then doubled again in $\sim$2.5 Gyr with \tfifty\ $= 5.08^{+0.05}_{-0.01}$ Gyr. Finally, the lookback time for 90\% of the mass to form is \tninety\ $= 1.18\pm0.01$ Gyr. We provide values of \tten, \ttwentyfive, \tfifty, and \tninety\ for both the PARSEC and BaSTI models in Table~\ref{tab:properties}. We discuss implications of the extended pause in star formation post-reionization in Section~\ref{sec:conclusions}.

\subsection{Comparison of NIRCam-based AMRs with Spectroscopic MDFs}\label{sec:amr}
As a check on the fits, we compare the best-fitting AMR solutions from the NIRCam data with trends based on spectroscopic abundances of stars of different ages and from gas-phase abundances measured from \hii\ regions in WLM. First, for a comparison of the AMR at intermediate ages, we use the metallicity distribution reported in \citet{Leaman2009} from 44 RGB stars that are spatially located in a similar region of the NIRCam data (labeled as the `bar' region by these authors). They find a mean metallicity [Fe/H] $= -1.14\pm0.04$ with an intrinsic spread of $\sigma=0.33$ dex based on converting summed equivalent widths of calcium II triplet lines from medium resolution (R $\sim$ 3400) spectroscopy to an [Fe/H] metallicity scale. This is slightly higher than the mean [Fe/H] value of $-1.28, \sigma=0.38$ dex found for a larger sample of 126 stars in WLM that cover the stellar disk out to larger radii \citep{Leaman2013}. The mean [Fe/H] $= -1.14$  can be compared with the [M/H]\footnote{Note that the spectroscopically measured metallicity corresponds to the [Fe/H], while the AMR from CMD-fitting is an estimate of the mean stellar metallicity quantity [M/H] that is used in stellar models.} range from $-1.2$ to $-0.8$ in the AMRs at intermediate times seen in Figure~\ref{fig:sfh_nircam}, which suggests general agreement between the spectroscopically-determined stellar metallicity distribution function and AMR trends. 

While a good general comparison, the AMRs are based on the composite stellar populations, whereas the spectroscopic MDFs were created from stars in a limited CMD region, namely stars that lie on the RGB and that are within $\sim$1 mag below the TRGB. Thus, to compare the AMRs with the MDFs more directly, we used the best-fitting SFH and AMR solutions to generate synthetic photometry for each stellar library. We then selected stars that lie on the RGB within $\sim1$ mag of the TRGB and calculated the mean [M/H] values and the standard deviations. Assuming a normal distribution, we find a $\langle$[M/H]$\rangle = -0.94; \sigma = 0.25$ dex for the BaSTI models and $\langle$[M/H]$\rangle = -0.92; \sigma = 0.22$ dex for the PARSEC models, in good agreement with the $\langle$[Fe/H]$\rangle = -1.14; \sigma = 0.33$ dex reported in \citet{Leaman2009}.

\begin{figure}
\begin{center}
\includegraphics[width=0.48\textwidth]{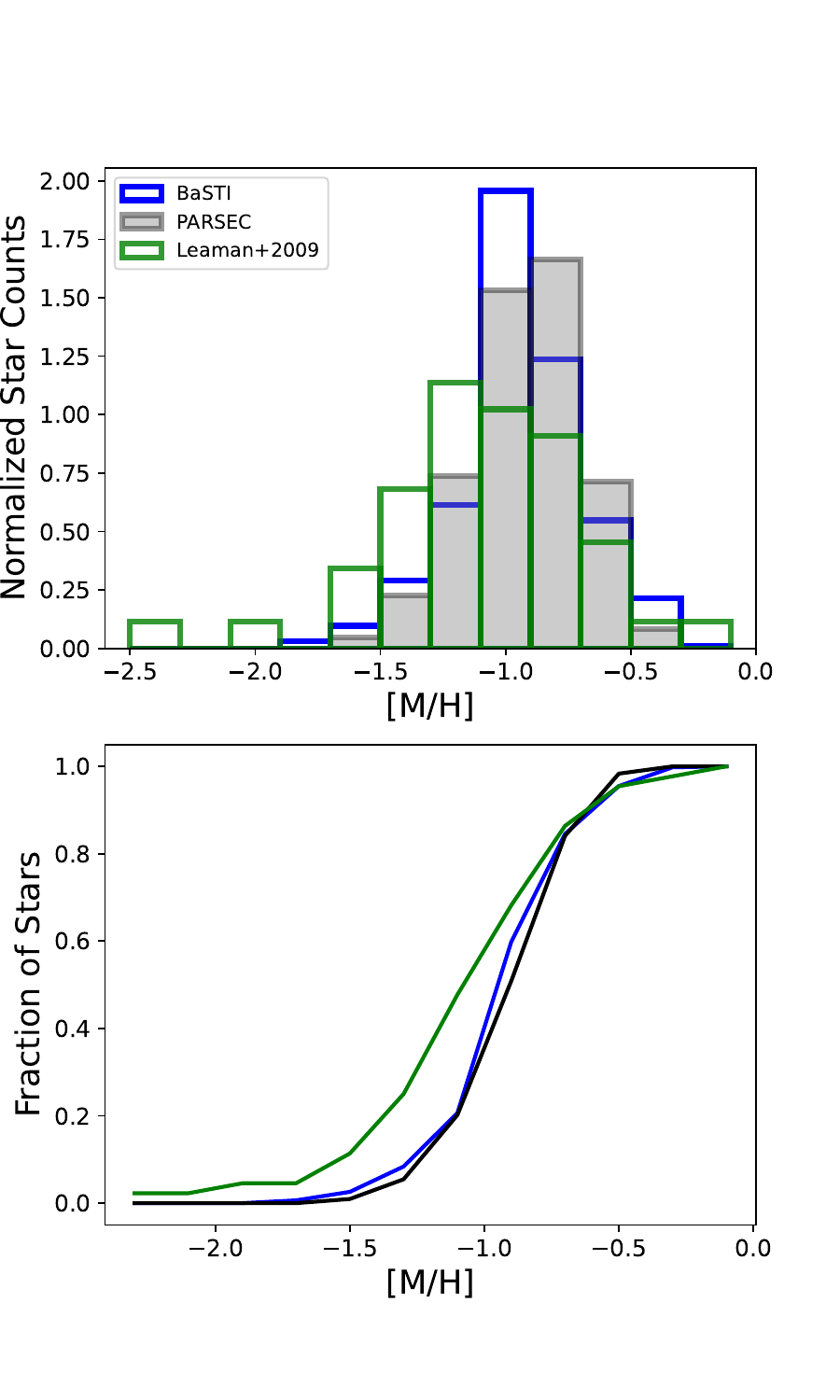}
\end{center}
\vspace{-0.45in}
\caption{Top panel: Comparison of the spectroscopically determined [Fe/H] MDF from \citet{Leaman2009} with the [M/H] MDFs estimated from the SFH and AMR derived with the PARSEC and BaSTI models, all normalized to total star counts. Bottom panel: The same information shown as cumulative distribution functions (CDFs). Based on 2-sample KS tests, the modelled CDFs have a high probability of being drawn from the spectroscopic CDF. See text for details.}
\label{fig:mdf_comparison}
\end{figure}

Figure~\ref{fig:mdf_comparison} shows that the [M/H] distributions (top panel) and cumulative distribution functions (bottom panel) from both libraries are consistent with the [Fe/H] values for the 44 stars from \citet{Leaman2009}. The largest difference is the presence of a metal-poor tail in the spectroscopically determined [Fe/H] values, which is absent in the AMR-based distribution functions. We note that the lack of a metal-poor tail in the modeled distributions is consistent with the corresponding lack of metal-poor blue horizontal branch stars in both the ACS and the NIRCam CMDs. A two-sided Kolmogorov–Smirnov (K–S) test performed on the cumulative metallicity distribution functions from the Leaman et al.\ data and our values showed there is 26\% probability of the results being drawn from the same distribution for the PARSEC models and 54\% for the BaSTI models; the difference in the K-S tests is likely driven by the different peaks and widths of the distributions, which is seen most clearly in the normalized histograms in the top plot. Regardless of the subtle differences, this is quite good agreement given the somewhat different spatial regions sampled, relatively smaller number of spectroscopic measurements, and the different techniques used in the comparison.

For a comparison of the AMR results at more recent times, we consider spectroscopic measurements of young stars and the present-day gas abundance. Specifically, the metallicity of 8 young A and B supergiant stars was reported to have a mean [Fe/H] of $-0.87\pm0.06$ \citep[][see also \citealt{Bresolin2006}]{Urbaneja2008}. These values are in very good agreement with the measured gas-phase oxygen abundance from \hii\ regions \citep[e.g., 12$+$log(O/H) $= 7.83\pm0.06$;][]{Lee2005}. We also note that there is evidence that WLM hosts young stars with higher metallicities based on the study of two supergiants \citep{Venn2003} and one massive O star \citep{Telford2021}. For comparison, the AMRs at the present day show [M/H] values ranging from $-0.6$ to $-0.8$, depending on the stellar library, which is again in good agreement with the spectra-based values. 

\begin{figure}
\begin{center}
\includegraphics[width=0.45\textwidth]{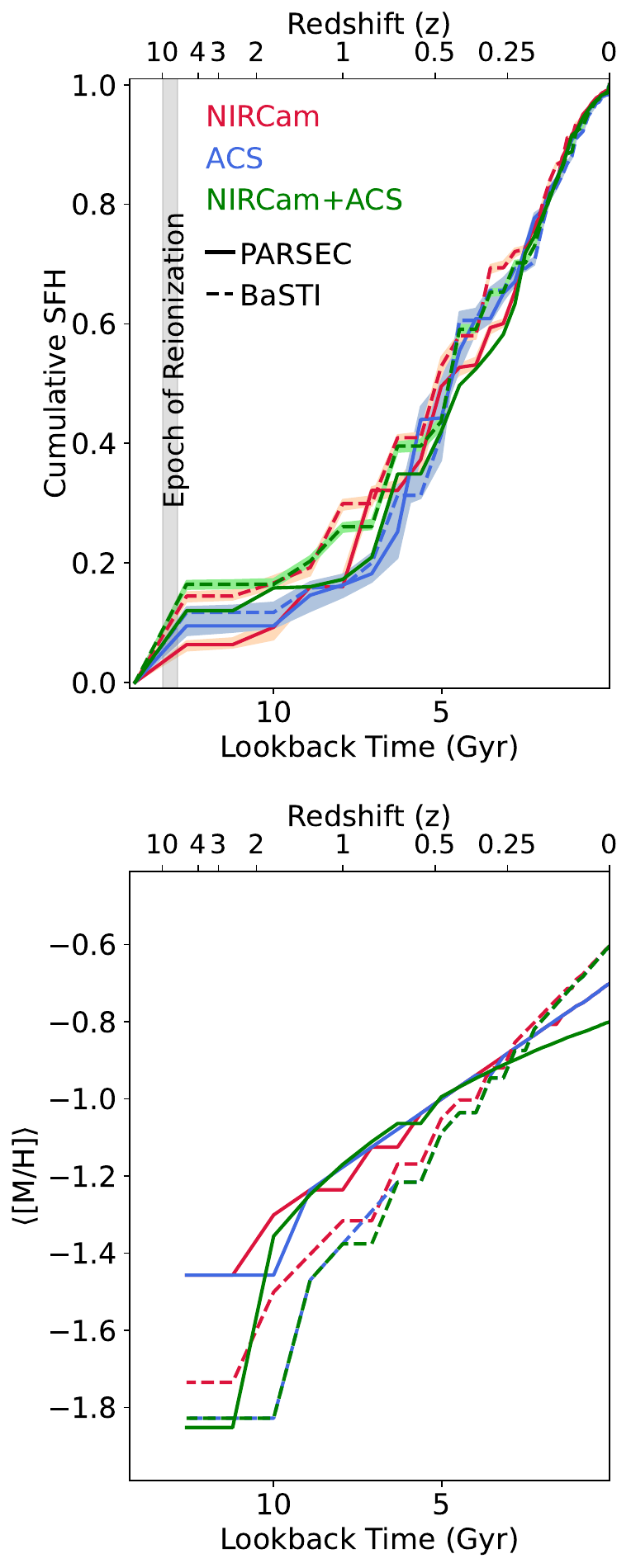}
\end{center}
\caption{SFHs (top) and AMRs (bottom) derived by fitting the NIRCam (red) and ACS (blue) stellar catalogs from the overlapping region using the PARSEC models (solid) and the BaSTI models (dashed). Also shown are the solutions derived by {\em simultanesouly} fitting the NIRCam and ACS CMDs (green). The shaded regions represent statistical uncertainties. The SFHs are all in excellent agreement.}
\label{fig:sfh_jwst_hst}
\end{figure}

\subsection{Comparison of NIRCam-based and ACS-based SFHs}\label{sec:sfh_jwst_hst}
Here, we focus on a quantitative comparison of the NIRCam- and ACS-based SFHs. The fits are based solely on the stellar catalogs from the overlapping footprint between NIRCam and ACS, as seen in Figure~\ref{fig:fov}, rather than the catalogs from the full field of view of either instrument. 

Figure~\ref{fig:sfh_jwst_hst} presents the SFH (top) and AMR (bottom) solutions from NIRCam data (red) and ACS data (blue) using the PARSEC (solid lines) and BaSTI (dashed lines) models. Shaded regions represent statistical uncertainties on the fits. The SFH results for NIRCam and ACS data from both the PARSEC and BaSTI libraries are in excellent agreement with each other at all lookback times. These results confirm the efficacy of the F090W - F150W filter combination and stellar evolution libraries in the near-infrared for deriving SFHs relative to the HST optical filters.

Figure~\ref{fig:sfh_jwst_hst} also shows the SFHs from the simultaneous fits to the NIRCam and ACS CMDs (green lines). These solutions are also in excellent agreement with the SFHs and AMRs derived independently from the NIRCam data and from the ACS data (i.e., single CMD-fitting). The agreement indicates that the SFHs recovered by fitting just one CMD in the optical or in the near-infrared are not significantly biased compared to solutions recovered from simultaneously fitting two CMDs that intrinsically contain more information, and demonstrates a degree of robustness for the single CMD fit.

As a further comparison of the different SFH fits, we calculate the total stellar mass recovered from the different data sets (NIRCam, ACS, NIRCAM+ACS) and stellar libraries (PARSEC, BaSTI) in the overlapping field of view. The values are in close agreement ranging from log(\mstar/\msun) of 6.54 to 6.65, providing additional evidence of the consistency between the SFH recovered from the two observatories. See Table~\ref{tab:properties} for individual values. 

Compared to the SFH results derived from the full NIRCam field of view (Figure~\ref{fig:sfh_nircam}), the SFHs from the overlapping NIRCam - ACS region (Figure~\ref{fig:sfh_jwst_hst}) show a slightly slower start to star formation. This suggests a lower median stellar age at larger radii. Similarly, the AMRs in the overlapping NIRCam - ACS data have slightly lower values than the AMRs from the full NIRCam field, suggesting lower mean metallicities at larger radii. 

\citet{Albers2019}, who reconstructed the SFH from the full ACS data set, also report the SFH from an outer field in WLM using deep HST data from the WFC3 instrument. Compared with the SFHs from regions imaged with the ACS and NIRCam, these results suggest a strong radial age gradient. Quantitatively, the inner fields formed 50\% of their stellar mass $\sim5$ Gyr ago compared with $\sim10$ Gyr in the outer field located only $\sim0.4$ kpc off the minor axis. Future work from our team will combine the JWST NIRCam field, the different HST fields, and the JWST NIRISS field observed as part of the ERS program to explore this in greater detail (Roger Cohen et al.\ in preparation). 

\section{Discussion and Summary}\label{sec:conclusions}
\subsection{On the Efficacy of NIRCam Imaging for SFH Work}
In this work, we use deep JWST NIRCam imaging of the Local Group star-forming galaxy, WLM, from the ERS-1334 program to derive the SFH and AMR of the central field in the galaxy using two sets of stellar libraries (PARSEC, BaSTI). We have leveraged existing optical imaging obtained with the HST ACS instrument on an overlapping region \citep{Albers2019} to perform a one-to-one comparison of NIRCam-based and ACS-based SFH results. We find the SFHs and AMRs derived independently and simultaneously fit using data from both great observatories to be in excellent agreement for the two stellar libraries. In addition, the NIRCam-based AMRs agree with spectroscopic metallicity distribution functions, providing an independent check on the recovered solutions. 

This first detailed look at a SFH using NIRCam data confirms the efficacy of recovering the SFH with the F090W - F150W filter combination relative to the workhorse optical filters on HST that have been used for SFH work for the past two decades. These results also provide validation of the sensitivity and accuracy of stellar evolution libraries in the infrared relative to the optical. With JWST's greater sensitivity and resolution, detailed and robust SFHs can be recovered with greater efficiency than has been possible with HST and for galaxies at greater distances. It also validates that future SFH results from JWST can be directly compared with existing SFH studies from HST data without complications of significant biases due to observations from these different observatories and wavelength coverage. 

\subsection{The Post-Reionization Pause in Star Formation in WLM}
To date, only a few gas-rich isolated dwarf irregular (dIrr) galaxies have robustly reconstructed ancient SFHs from deep CMDs in the literature. In order of decreasing mass, these include IC~1613 \citep{Skillman2003},  WLM \citep[][and the present work]{Albers2019}, Leo~A \citep{Cole2007}, and Aquarius \citep{Cole2014}. While the most massive of these galaxies (IC1613; \mstar\ $=1\times10^8$ \msun) shows relatively constant star formation across cosmic time, the remaining three galaxies (\mstar\ $\sim2\times10^6$ to $4\times10^7$ \msun) exhibit some level of early star formation followed by a pause that occurs post-reionization. These galaxies hosted a similarly early onset of star formation but then a temporary quiescent period ensues as, presumably, the metal-poor gas present in and/or accreted onto the galaxies was prevented from cooling efficiently, with myriad heating sources (e.g., cosmological reionization, the local far-UV radiation field, heating from internal stellar feedback, etc.) potentially playing important roles.

Although this is a very small sample, the emerging pattern is intriguing, suggesting that even in the stellar mass range of $10^6-10^7$ \msun, galaxies experience a post-reionization pause in star formation. Similarly deep CMDs of additional {\em isolated} dIrr galaxies are needed to confirm and further investigate this trend. JWST is currently the only facility with the capabilities to increase the sample size of isolated galaxies (i.e., galaxies outside the Local Group) with robust ancient SFHs. 

\begin{acknowledgments}
This work is based in part on observations made with the NASA/ESA/CSA James Webb Space Telescope. The data were obtained from the Mikulski Archive for Space Telescopes at the Space Telescope Science Institute, which is operated by the Association of Universities for Research in Astronomy, Inc., under NASA contract NAS 5-03127 for JWST. These observations are associated with program ERS-1334. This research is also based in part on observations made with the NASA/ESA Hubble Space Telescope obtained from the Space Telescope Science Institute, which is operated by the Association of Universities for Research in Astronomy, Inc., under NASA contract NAS 5–26555. These observations are associated with program HST-GO-13678. Support for this work was provided by NASA through grants ERS-1334 from the Space Telescope Science Institute, which is operated by AURA, Inc., under NASA contract NAS5-26555. This research has made use of NASA Astrophysics Data System Bibliographic Services and the NASA/IPAC Extragalactic Database (NED), which is operated by the Jet Propulsion Laboratory, California Institute of Technology, under contract with the National Aeronautics and Space Administration. LG and AM acknowledge financial support from Padova University, Department of Physics and Astronomy Research Project 2021 (PRD 2021).
\end{acknowledgments}

\facilities{JWST, Hubble Space Telescope}

\software{This research made use of Astropy,\footnote{http://www.astropy.org} a community-developed core Python package for Astronomy \citep{astropy:2013, astropy:2018, astropy:2022}; {\tt DOLPHOT} \citep{Dolphin2000, Dolphin2016}; and \textsc{match} \citep{Dolphin2002, Dolphin2013}.}

\begin{figure*}
\gridline{\fig{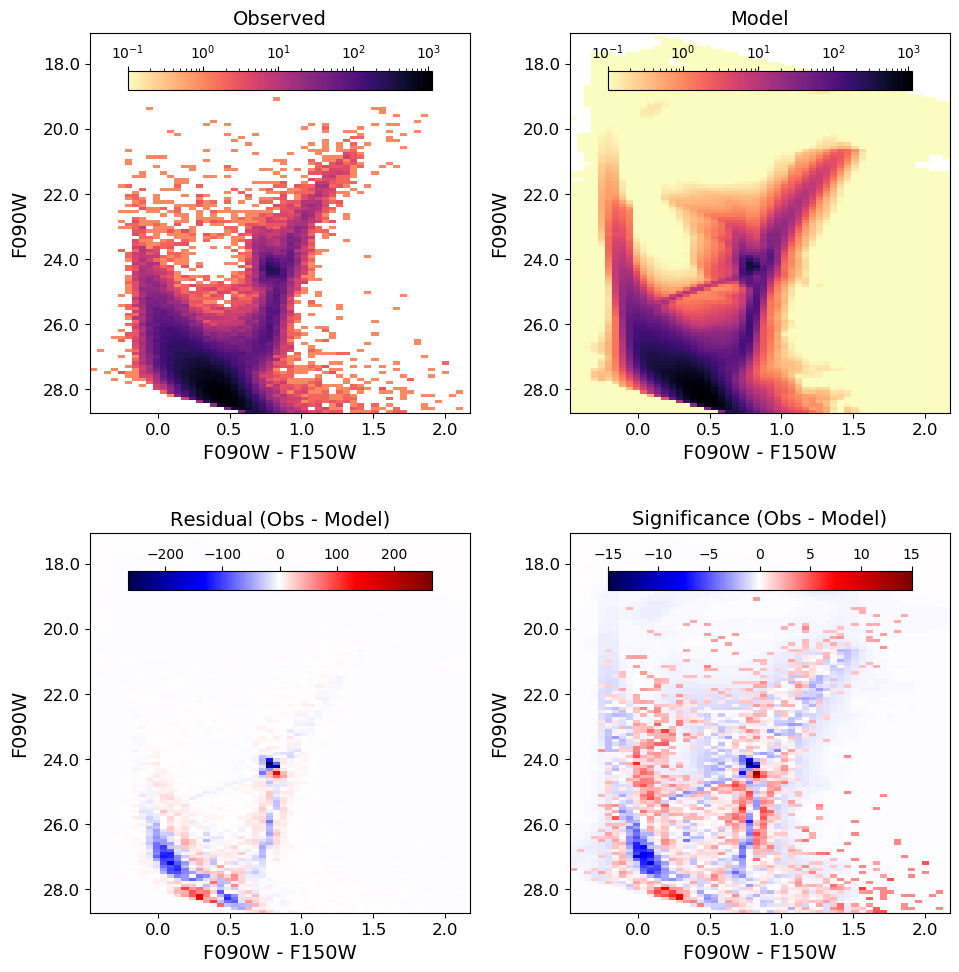}{0.45\textwidth}{(a) JWST NIRCam data fit with the PARSEC library}
    \fig{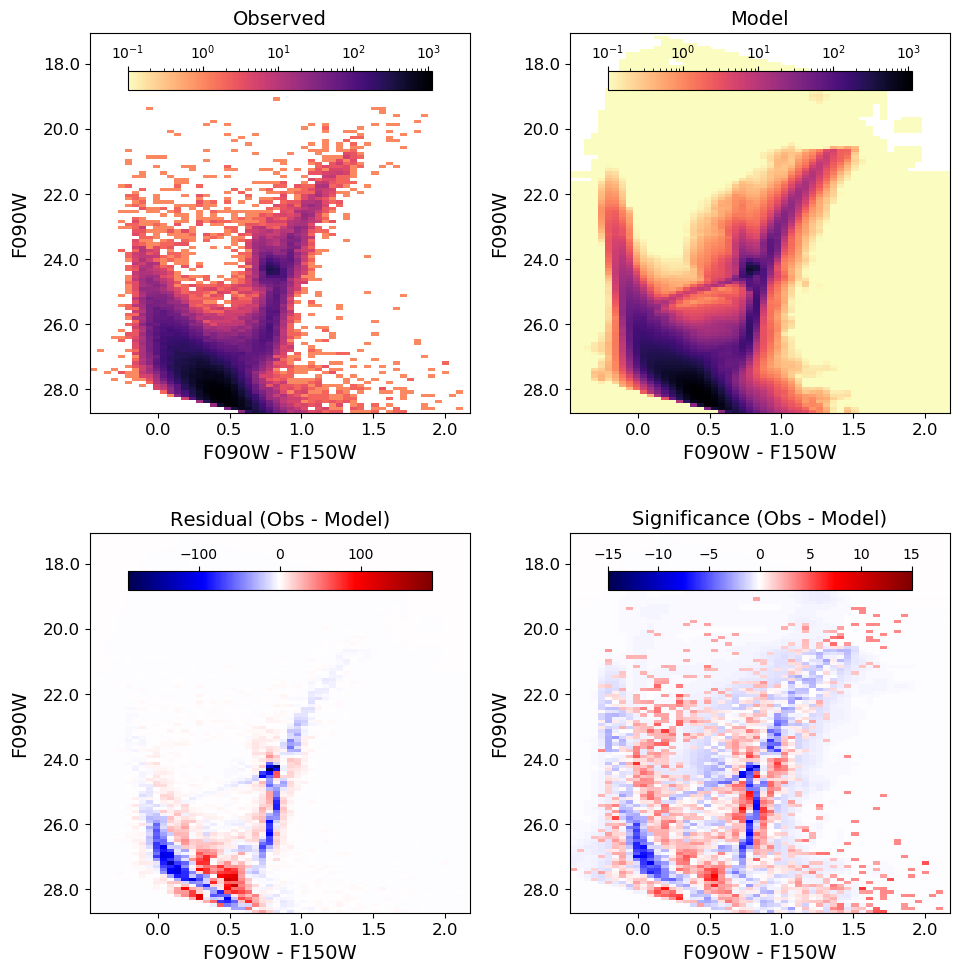}{0.45\textwidth}{b) JWST NIRCam data fit with the BaSTI library}}
\gridline{\fig{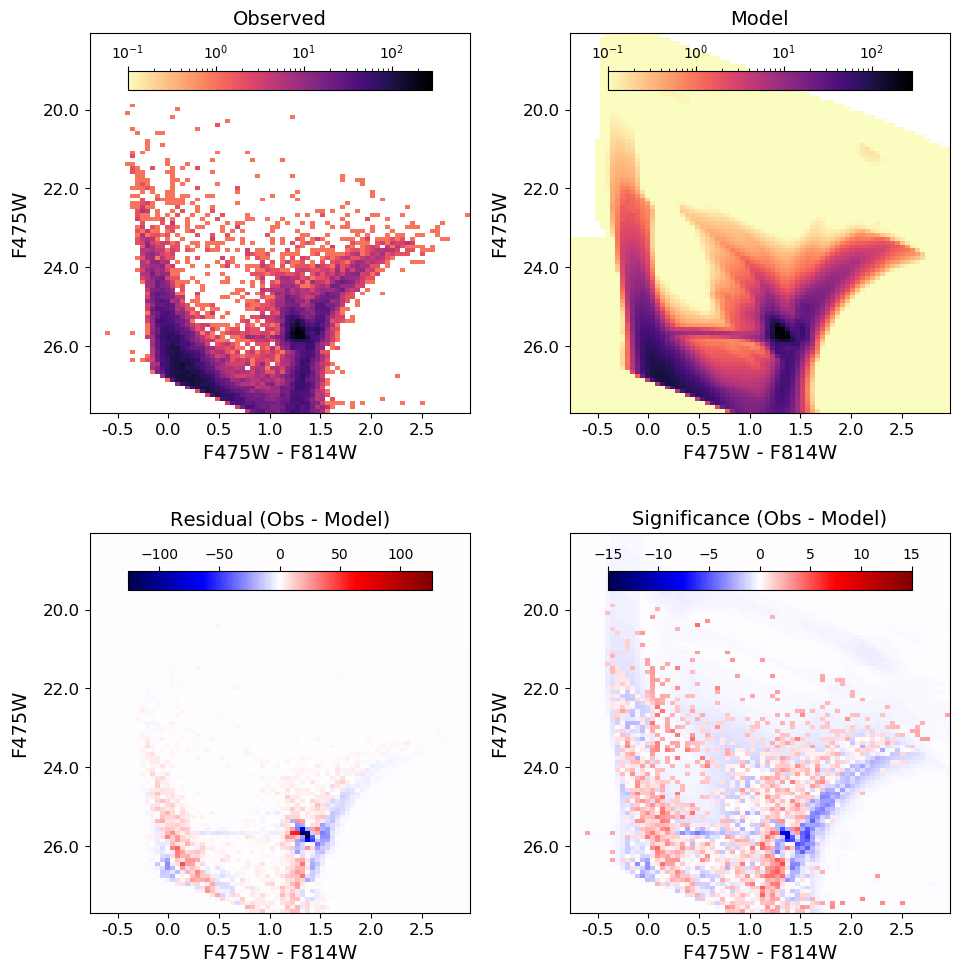}{0.45\textwidth}{(c) HST ACS data fit with the PARSEC library}
     \fig{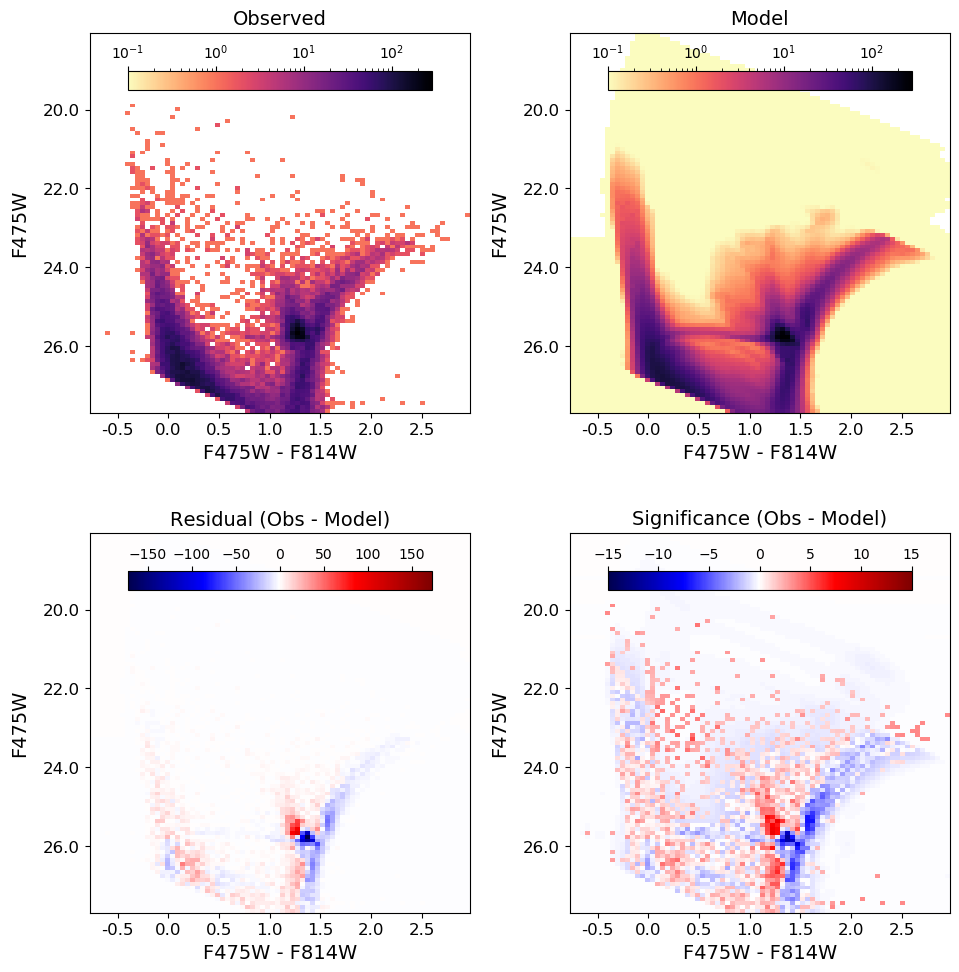}{0.45\textwidth}{(d) HST ACS data fit with the BaSTI library}}
\caption{Results from the simultaneous fits to the NIRCam (subplots a, b) and ACS (subplots c, d) stellar catalogs in the common field of view. The figure format is the same as shown in Figure~\ref{fig:residuals}.}
\label{fig:residuals_2cmds}
\end{figure*}

\appendix\label{sec:app}
For completeness, in Figure~\ref{fig:residuals_2cmds} we present the results from simultaneously fitting the NIRCam and ACS CMDs. The figure has the same format as Figure~\ref{fig:residuals}: the top panels show the fits to the NIRCam data, the bottom panels show the fits to the ACS data, with the PARSEC models (left) and the BaSTI models (right). The patterns between the observed and modelled CMDs seen in residual significance plots (lower right panels in each sub-plot) are quite similar to those from the single CMD fits shown in Figure~\ref{fig:residuals}. 

\renewcommand\bibname{{References}}
\bibliographystyle{apj}
\bibliography{ms_final.bbl}

\end{document}